\documentclass[sigconf]{acmart}
\AtBeginDocument{%
  \providecommand\BibTeX{{%
    \normalfont B\kern-0.5em{\scshape i\kern-0.25em b}\kern-0.8em\TeX}}}

\AtBeginDocument{%
  \RequirePackage{hyperxmp}
}
\pdfoutput=1 

\setcopyright{acmcopyright}
\copyrightyear{2018}
\acmYear{2018}
\acmDOI{XXXXXXX.XXXXXXX}

\acmConference[Conference acronym 'XX]{Make sure to enter the correct
  conference title from your rights confirmation emai}{June 03--05,
  2018}{Woodstock, NY}
\acmBooktitle{Woodstock '18: ACM Symposium on Neural Gaze Detection,
 June 03--05, 2018, Woodstock, NY} 
\acmPrice{15.00}
\acmISBN{978-1-4503-XXXX-X/18/06}

\usepackage{xspace}
\usepackage{subcaption}
\usepackage{booktabs}
\usepackage{colortbl}
\usepackage{tikz}

\newcommand{\lightbluecircled}[1]{%
  \tikz[baseline=(char.base)]{
    \node[shape=circle,draw={rgb,255:red,172;green,203;blue,236},fill={rgb,255:red,172;green,203;blue,236},inner sep=1pt,text=white] (char) {\small\bfseries #1};
  }%
}

\newcommand{\lightpurplecircled}[1]{%
  \tikz[baseline=(char.base)]{
    \node[shape=circle,draw={rgb,255:red,216;green,192;blue,216},fill={rgb,255:red,216;green,192;blue,216},inner sep=1pt,text=white] (char) {\small\bfseries #1};
  }%
}

\newcommand{\lightgreencircled}[1]{%
  \tikz[baseline=(char.base)]{
    \node[shape=circle,draw={rgb,255:red,218;green,229;blue,179},fill={rgb,255:red,218;green,229;blue,179},inner sep=1pt,text=white] (char) {\small\bfseries #1};
  }%
}

\definecolor{matilda}{HTML}{DADDB3}
\definecolor{lavender}{HTML}{D8C0D8}
\definecolor{bruce}{HTML}{ACCBEC}

\newcommand{\system}{DraftMarks\xspace} 
\newcommand{\systems}{DraftMarks'\xspace}

\begin{document}



\title[Transparency in Human-AI Co-Writing Through Skeuomprohic Process Traces]{\system: Enhancing Transparency in Human-AI Co-Writing Through Interactive Skeuomorphic Process Traces}


\author{Momin N. Siddiqui}
 \affiliation{%
  \institution{Georgia Institute of Technology}
   \country{USA}
 }
 \email{msiddiqui66@gatech.edu}

 \author{Nikki Nasseri}
 \affiliation{%
  \institution{University of California, Berkeley}
   \country{USA}
 }
 \email{nassern1@berkeley.edu}
 
\author{Adam Coscia}
\orcid{0000-0002-0429-9295}
\affiliation{%
  \institution{Georgia Institute of Technology}
  \city{Atlanta}
  \state{Georgia}
  \country{USA}
}%
\email{acoscia6@gatech.edu}

 \author{Roy Pea}
 \affiliation{%
  \institution{Stanford University}
   \country{USA}
 }
 \email{roypea@stanford.edu}

\author{Hari Subramonyam}
 \affiliation{%
  \institution{Stanford University}
   \country{USA}
 }
 \email{harihars@stanford.edu}

\renewcommand{\shortauthors}{Siddiqui et al.}

\begin{abstract}
As generative AI becomes part of everyday writing, questions of transparency and productive human effort are increasingly important. Educators, reviewers, and readers want to understand how AI shaped the process. Where was human effort focused? What role did AI play in the creation of the work? How did the interaction unfold? Existing approaches often reduce these dynamics to summary metrics or simplified provenance. We introduce \system, an augmented reading tool that surfaces the human–AI writing process through familiar physical metaphors. \system employs skeuomorphic encodings such as eraser crumbs to convey the intensity of revision, and masking tape or smudges to mark AI-generated content, simulating the process within the final written artifact. By using data from writer-AI interactions, \systems algorithm computes various collaboration metrics and writing traces. Through a formative study, we identified computational logic for different readership, and evaluated  \system for its effectiveness in assessing AI co-authored writing. 
\end{abstract}

\keywords{User interface, Human-AI collaboration, Writing analytics, Skeuomorphism, Large language models}

\begin{teaserfigure}
  \includegraphics[width=\textwidth]{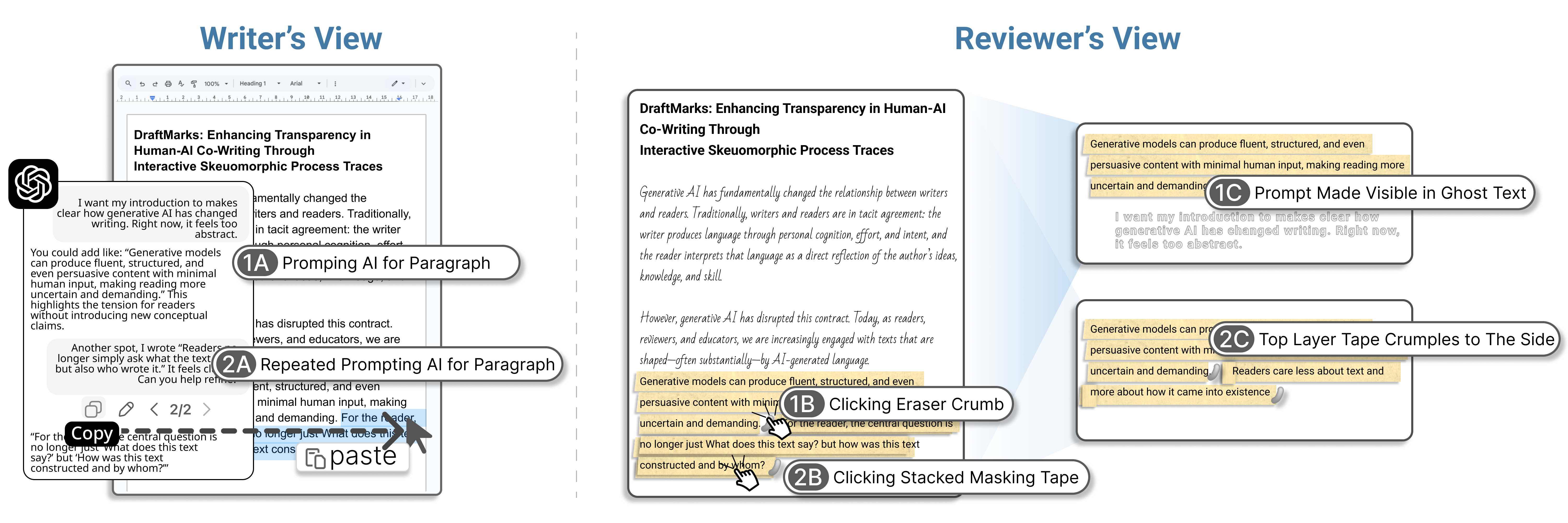}
  \caption{\system reveals the hidden dynamics of human–AI collaboration to readers. The Writer's View shows AI generation calls: single prompts (1A) and repeated prompting (2A). The Reviewer's View surfaces these interactions within the final text through skeuomorphic design: eraser crumb and ghost text (1B-1C), and stacked masking tape (2B-2C).}
  \label{fig:teaser}
\end{teaserfigure}


\maketitle

\section{Introduction}

Generative AI has fundamentally changed the relationship between writers and readers. Traditionally, writers and readers are in tacit agreement: the writer produces language through \textit{personal} cognition, effort, and intent, and the reader interprets that language as a direct reflection of the author's ideas, knowledge, and skill~\cite{hayes2016identifying,scardamalia1987knowledge}. Whether reading an essay, an email, or a scientific article such as this article, the reader makes inferences about the author's expertise, creativity, and trustworthiness based on the text itself~\cite{Miao2023EthicalDI}. However, generative AI has disrupted this contract. Today, as readers, reviewers, and educators, we are increasingly engaged with texts that are shaped---\textit{often substantially}---by AI-generated language. Generative models can produce fluent, structured, and even persuasive content with minimal human input, making reading more uncertain and demanding. For the readers, the central question is no longer just \textit{What does this text say?} but also \textit{How was this text constructed and by whom?}

Unfortunately, readers lack effective tools and transparent design affordances to make these distinctions and critically assess texts co-authored with AI as a writing partner. Current approaches are primarily limited to AI usage disclosures~\cite{Cho2023PaperCardFR}, sharing AI prompts and conversation logs~\cite{Yoo2023RECIPEHTA}, and token-level attribution maps or summary visualizations of AI contributions~\cite{Shibani:2023:HumanMachineWriting, Hoque:2024:HaLLMark}. However, these approaches are often external to the text rather than \textit{within} it, requiring the reader to step out of the reading experience to consult metadata or interpret visualizations detached from the comprehension flow. This separation becomes problematic when readers are making local interpretive decisions: \textit{Is this sentence the author's reasoning or the model's default phrasing? Was this claim carefully revised or copied verbatim?} For example, knowing that a paragraph was iteratively revised might lead a reader to trust its reasoning more; recognizing that a definition was injected via AI may prompt the reader to cross-check its accuracy. These decisions are part of how the readers build mental models about what matters in writing, what to question critically, and what reflects human expertise~\cite{Shibani:2024:CriticalInteractionAI}. When process information is abstracted away, the reader's comprehension remains incomplete, obscuring the \textit{cognitive labor} --- or lack thereof --- behind the text.

In this work, we explore the use of skeuomorphic design~\cite{Coyne:1992:MetaphorDesign, Blackwell:2006:ReificationMetaphorDesignTool,Norman:2014:DesignEverdayThings}, i.e., visual encodings grounded in familiar physical metaphors, to \textit{surface} human-AI writing interactions in context. Rather than abstract icons or disconnected dashboards, skeuomorphic cues such as masking tape, eraser crumbs, smudges, and scribbled margins could provide legible embedded markers of writing activity. Akin to traditional settings, readers may draw meaning about authorial effort from crossed-out sentences, margin notes, use of white-out tape, or layered corrections of pen over pencil (i.e., physical traces into the writer's thinking). Readers rely on physical signals to support \textit{inline} interpretation of AI coauthored text. These signals should indicate where the writer invested effort, where they accepted or rejected AI suggestions, and where the text evolved through iteration. For example, a masking tape overlay might indicate a passage written primarily by an AI model, while eraser marks could show a region of intensive human-AI iteration. These metaphor-driven encodings draw on readers' pre-existing literacies for interpreting physical documents and editing marks, repurposing them for AI-mediated writing.

In our technique, we formalize this approach by linking interaction logs of human-AI writing to skeuomorphic visualizations that appear within the text document. \system follows a model-view-controller paradigm, where the model processes interaction data to compute metrics such as contribution ratios, edit frequency, and prompt iteration depth, and the view maps them to visual overlays that retain the spatial and rhetorical coherence of the original text. To devise the controller logic that determines how and when these visualizations appear, we conducted a formative study with 21 readers across three readership roles (teachers, academic reviewers, and general readers), gathering insights on their interpretation needs and preferences for process transparency. We found that while teachers prefer detailed traces to support formative assessment of student learning, academic reviewers prefer minimal process information that prioritizes intellectual contribution over process details. General readers seek authenticity markers to evaluate author effort and trustworthiness. These findings informed our controller design, which adapts the same underlying interaction data into stakeholder-specific visualizations that balance comprehension support with evaluative needs.

As show in Figure~\ref{fig:teaser}, the human-AI collaboration trace (1A and 2A) is rendered using \system to surface co-writing insights directly into the text. In case of the reviewer view, we see single masking tape and its variant of stacked masking tape, encoding single prompt and iterative prompting respectively. Together, these features in \system enable \textit{process-aware reading} by offering \textit{transparency} into how the text was constructed with AI. A between-subject evaluation with 70 participants split between \system and a baseline comprising of collaboration chat history and final essay artifact in split view revealed that participants in \system condition had better comprehension of writing process and also reported higher self-reported transparency in the co-writing process. 

In summary, our contributions are:
\begin{itemize}
    \item \system~--- a visualization technique that uses skeuomorphic design to embed writing process signals directly within AI-coauthored text.
    \item A tool implementation following the model-view-controller paradigm that takes human-AI collaboration data from writing sessions and automatically generates \system visualizations, mapping interaction data to in-text visual cues.
    \item A design probe with 21 readers across 3 reader types (educators, reviewers, and general readers) that informed the design of our controller logic and revealed key requirements for process transparency in AI-coauthored text.
    \item A user evaluation showing that \system supports more accurate and transparent interpretations of the co-writing process than a baseline disclosure condition, while maintaining usability and low cognitive load.
\end{itemize}

\section{Related Work}
\label{sec:related_work}
AI co-authorship, the practice of collaboratively producing text with large language models, has introduced a fundamental change in how we think about writing, authorship, and intellectual ownership~\cite{lee2024design,lee2022coauthor}. These ``co-creative systems'' operate along a continuum between user and system contributions, blurring clear lines of attribution and complicating questions of responsibility and credit~\cite{Buschek2021NinePP}. Academic institutions and publishers have responded with caution: top-tier journals, including \textit{Nature} and \textit{Science}, prohibit listing AI as an author and require clear disclosure of AI participation~\cite{Sohail2023DecodingCA, ACMPB:2023:PolicyAuthorship,AG:2023:AGNewClauses}. However, cases of AI systems such as ChatGPT~\cite{chatgpt} being listed as authors have already appeared in academic databases~\cite{Bahammam2023AdaptingTT}. 

Central to these concerns is a question of accountability. Tools such as ChatGPT ``cannot be held responsible'' explaining why editors and publishers reject AI coauthorship outright~\cite{Parikh2023ChatGPTPreliminaryOW}. Furthermore, AI tools introduce an ``AI Ghostwriter Effect'' where users do not perceive ownership of AI-generated text, but also refrain from publicly acknowledging AI's role~\cite{Draxler2023TheAG}. These tensions are particularly acute in educational settings, where learning to write is as important (if not more so) than the final product. Several studies note that reliance on AI-generated writing can erode students' critical thinking, expressive capabilities, and revision skills~\cite{Rane2024EnhancingTQ}. Teachers often express concern that students will bypass the cognitive work of structuring ideas or developing arguments, a shortcut that undermines learning goals and the validity of writing as an assessment~\cite{liu2024investigating}. Moreover, the growing presence of fluent but untraceable AI-generated prose makes it difficult for instructors to distinguish genuine student effort from automated composition~\cite{Najjar2025LeveragingEA,Jarrah2023UsingCI}. Similar concerns arise in peer review journal article submissions: reviewers may unknowingly assess work with extensive AI involvement, yet be unable to evaluate the quality of argumentation, originality, or authorial voice~\cite{Miao2023EthicalDI}.

To address this gap between private awareness and public disclosure, our work examines how visualizing the writing process can support more transparent, reflective, and interpretable human-AI collaboration. This section reviews previous work on LLM writing support tools, visualization systems to understand human-AI interaction, and text visualization techniques to analyze collaborative writing to motivate the design of \textit{\system}.

\subsection{LLM Writing Support Tools}
\label{sec:related_ai_writing_tools}

The ubiquity of LLM conversational interfaces has spawned a new paradigm of AI-assisted writing. Recent LLM-driven writing tools are increasingly aiming for a more synergistic human-AI partnership. This development is consistent with Douglas Engelbart's advocacy of the idea of augmenting human intellect through the coevolution of human and tool systems \cite{engelbart2021augmenting, engelbart1999bootstrapping}. He argued that humanity's ability to solve problems improves as our ``tool systems'' (our technology) and ``human systems'' (our organization, language, and methods) evolve together in a positive feedback loop. Tools such as Script\&Shift~\cite{Siddiqui:2025:ScriptShift}, VISAR~\cite{Zhang:2023:VISAR}, and CoAuthor~\cite{lee2022coauthor} focus on enabling writers to meaningfully integrate AI contributions within their own workflow, fostering a more interactive and collaborative process rather than the simple acceptance of suggestions.

Critically, most previous tools are mixed-initiative ~\cite{Zhang:2023:VISAR}, in which the LLM often takes a proactive role in influencing or even driving the writing process.
However, recent studies have raised ethical concerns about AI-assisted writing.
The use of AI assistance in writing raises significant concerns about decreased author agency and ownership~\cite{Mieczkowski:2022:AgencyExpertiseAI}, the generation of lackluster or stereotyped content~\cite{Chen:2024:AutoSpark, Gero:2022:Sparks}, and ethical issues, including plagiarism and the possibility for the views of the user of a tool to be unduly influenced~\cite{Weidinger:2021:EthicalSocialRisksLLMs,Jakesch:2023:CowritingLLMsUserViews}.
Although maintaining some mixed-initiative capabilities, AI-assisted writing tools often lend themselves to AI heavily influencing or even driving the writing process, potentially reducing the human to a more passive role~\cite{Mieczkowski:2022:AgencyExpertiseAI}, thus simply a 'human in the loop' rather than a 'human in the center of the loop'.
In addition, specific concerns have been raised about the use of AI by students in academic writing~\cite{DAgostino:2023:AcademicExperts, CTE:2024:EthicalUseAI,Bowman:2022:ChatbotHomework,Miao2023EthicalDI}.

To address these concerns, a critical question persists: How might we enable transparency in this collaborative dynamic? There is a pressing need to provide writers, educators, reviewers, and other stakeholders with clear and informative data and associated tools to enable detailing how AI was used during the writing process. As highlighted in~\citet{Hoque:2024:HaLLMark}, tools are needed to help writers track their agency and AI usage, particularly as institutions and publishers establish policies that require transparency in AI-assisted writing~\cite{ACMPB:2023:PolicyAuthorship,Park2023AuthorshipPO}.
This need for transparency in the process and insightful analytics motivates our development of \system.

\subsection{Visual Analytics For Human-AI Collaboration}
\label{sec:related_vis_systems}
Collaborative writing has a long history of research into how humans coordinate with other humans to accomplish writing tasks \cite{Birnholtz:2012:CollaborativeWriting}.
We draw inspiration from prior works that visualize contributions from multiple authors towards analyzing writing patterns.
Many prior tools focus on "distant reading," revealing aggregate patterns, structures, authorship contributions, or timelines without showing the fine-grained evolution of the text itself~\cite{Chen:2025:ComparingSpeakerBehavior, McNely:2012:AnalyticsCollabWriting,Kim:2024:PromptAnalytics, Wang:2015:DocuViz, Viegas:2004:HistoryFlowVis}.
For example, DocuViz \cite{Wang:2015:DocuViz} is a tool for analyzing human-human collaborative writing post hoc.
These visualizations can be effective for understanding contribution ratios or identifying cooperation and conflict patterns, but are often difficult for laypeople to interpret, and can obscure the close reading of content changes~\cite{Viegas:2004:HistoryFlowVis}.
Techniques like highlighting authorship~\cite{Hoque:2024:HaLLMark,Torres:2019:VisAuthorship,Chevalier:2010:TextAnimatedTransitions} struggle to represent the complex history of revisions, especially when content ownership changes.
In contrast, fewer tools support "close reading" of collaborative revisions.
Some approaches use text animations or highlighting within a timeline view to show how content changes over time~\cite{Chevalier:2010:TextAnimatedTransitions}.
Other efforts have explored sentence-level visualizations~\cite{Shibani:2023:HumanMachineWriting}, but these tend toward abstract encodings that hinder close reading and may not capture metrics such as iteration depth or the nature of writer engagement.
Although useful, these tools often focus mainly on the \textit{what} of the substance of the changes, not necessarily on the \textit{ how} of the efforts of the writing processes underlying text production.

Developing visual analytics systems for human-AI collaboration is a relatively nascent but growing area of research.
Wang et al. describe the characteristics of systems that foster human-AI collaboration, which can be explored and analyzed using visual analytics systems \cite{Wang:2020:HumanHumanToHumanAI}.
Rogers and Crisan develop a methodology for visualizing collaborative human-AI processes in data work \cite{Rogers:2023:HumanAICollabVis}.
Li et al. characterize the collaborative roles of humans in a human-AI storytelling workflow, which can be visualized \cite{Li:2024:DataStorytellingHumanAI}.
Several domains have been the focus of visual analytics systems that facilitate human-AI collaboration, mainly in the area of interactive data analysis \cite{Sperrle:2021:AIGuidedTopicModeling}.
For example, StuGPTViz is a visual analytics system for post hoc analysis of student-LLM collaboration in education tasks \cite{Chen:2025:StuGPTViz}.
Graphologue \cite{Jiang:2023:Graphologue} and Sensecape \cite{Suh:2023:Sensecape} are systems that facilitate interactive human-AI collaboration on sensemaking tasks.
There is a recognized need for visualizations that specifically illuminate critical engagement with AI suggestions across domains~\cite{Shibani:2024:CriticalInteractionAI}.

Yet while there is prior work on visualizing human-AI collaboration, visual analytics tools specifically for gaining insights into human-AI collaborative writing are almost nonexistent.
However, HaLLMark \cite{Hoque:2024:HaLLMark} is a visual analytics system that represents a significant step, offering interactive visualizations to support provenance tracking and transparency regarding AI use, with the aim of helping writers maintain agency and comply with policies~\cite{ACMPB:2023:PolicyAuthorship, USCO:2023:CopyrightRegistration, ACMPB:2023:PolicyAuthorship}.
It captures key metrics such as prompt categories and AI response integration.
However, the authors acknowledge its limitations for analyzing long-form writing and suggest future work towards platform-independent tracking.
Crucially, HaLLMark focuses primarily on provenance (what text came from where) and summary statistics, offering less insight into the productive struggle or iterative refinement process undertaken by the writer, aspects often of interest to readers, educators, and reviewers. The productive struggle of a learner is of special interest to educators, as it signals the participation of cognitive processes associated with advances in learning \cite{Bjork2011making}.
Thus, we identify a critical gap: the need for visualizations that go beyond simple contribution ratios and provenance specifically in writing tasks.
\system~ aims to fill this gap by visualizing more complex process metrics reflecting the writer's effort, iteration, and metacognitive engagement with the AI, directly within the context of the final document.

\subsection{Text Visualization Techniques}
\label{sec:related_vis_techniques}

Text visualization encompasses a wide range of techniques for representing textual data~\cite{Brath2021:VisualizingWithText,Viegas:2008:TagClouds,Viegas:2009:ParticipatoryVisWordle,Wattenberg:2008:WordTree, Wise:1995:GalaxiesThemeScapes}.
Common methods include \textit{word clouds} to summarize the frequency of words~\cite{Viegas:2008:TagClouds} and \textit{tree structures} to show concordance~\cite{Wattenberg:2008:WordTree}.
Many techniques focus on ``distant reading,'' analyzing large corpora to reveal patterns, topics, and relationships through spatial layouts, network diagrams, or temporal flows~\cite{vanHam:2009:PhraseNets,Gad:2015:ThemeDelta}.
Other techniques support "close reading" by overlaying information on the text itself, such as highlighting phrases based on the underlying data or tags \cite{Chevalier:2010:TextAnimatedTransitions}.
Recent work has also explored using AI and visualization to analyze specific textual features, such as character representation or sonic properties in poetry~\cite{Nguyen:2016:SensePath}, or to visualize the structure of LLM responses themselves~\cite{Jiang:2023:Graphologue,Suh:2023:Sensecape}.
Our work draws on this rich history, but focuses specifically on embedding visualizations of the writing process \textit{within} the document to support interpretation of human-AI collaboration during close reading. This strategy necessitates moving beyond standard text visualization techniques towards metaphors that convey process history, leading us to explore skeuomorphism, as we describe in the next section.

\begin{figure*}
    \centering
    \includegraphics[width=\linewidth]{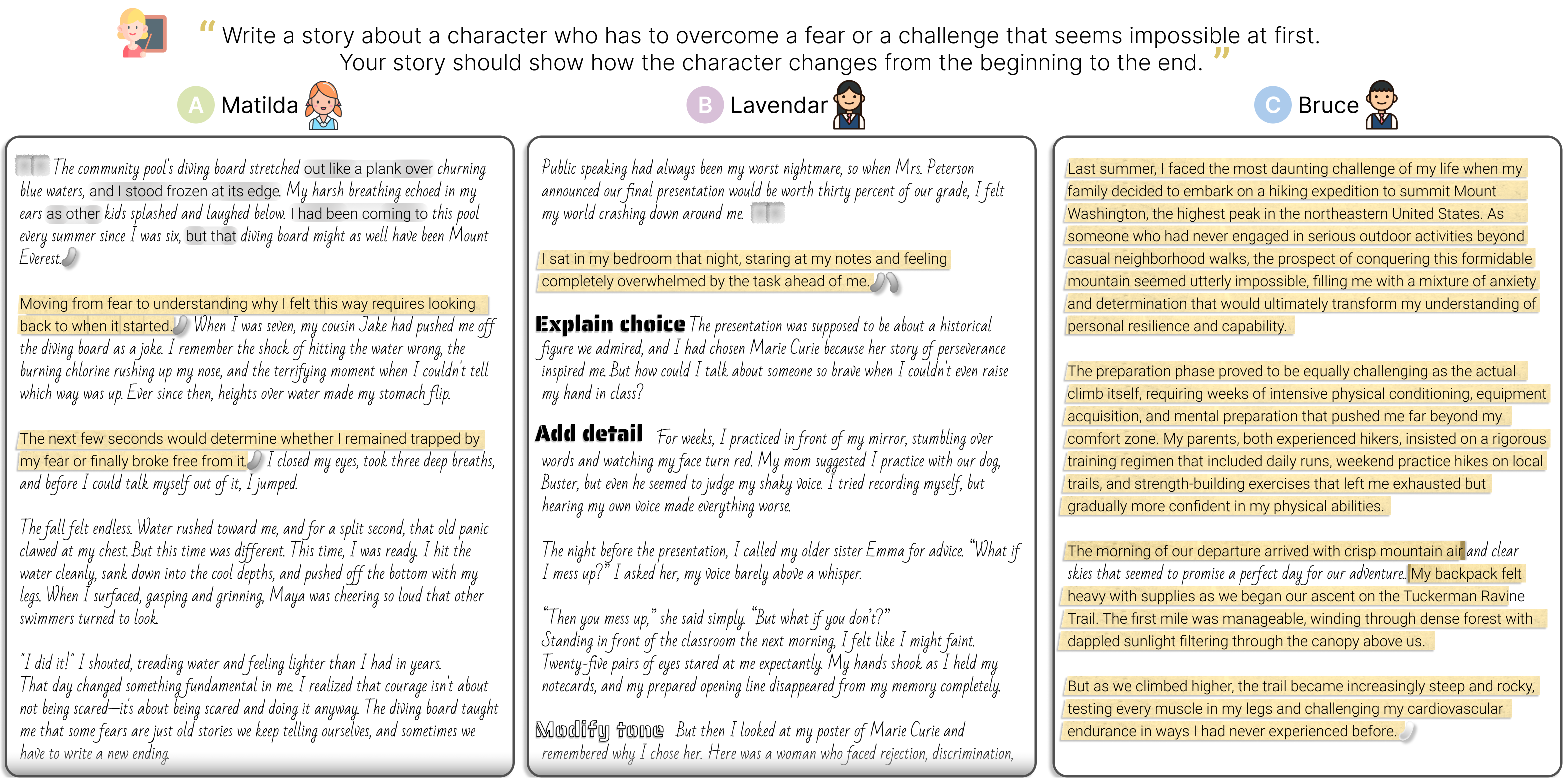}
    \caption{Human-AI Co-Writing Essay Samples.}
    \label{fig:all_essay}
\end{figure*}

\section{User Scenario}

At a high level, \system is a web-based \textit{interactive text viewer} designed to make the invisible dynamics of human-AI writing legible to readers by embedding process cues directly into the text. As shown in Figure~\ref{fig:all_essay}, the interface overlays skeuomorphic encodings---visual marks inspired by everyday writing materials---to reveal how a piece of text came to be. The augmentations, including masking tape, eraser crumbs, smudge marks, ghost text, stencil marks, and residual glue, were selected for their familiarity and semantic fit with the underlying activity they represent, such as provenance, erasure, or absence (discussed in more detail in Section~\ref{sec:view}). This visual language transforms the final draft into a layered record of interaction, preserving readability while revealing the writing processes. 

To understand the user experience of \system, let us consider an example scenario of Miss Jones, an eighth-grade English teacher. In her classroom, students are increasingly experimenting with generative AI to scaffold their narrative writing. Miss Jones has established a clear classroom policy: \textit{``You are allowed to use AI for brainstorming and improving tone. However, you must generate your own original ideas and may not use AI to write complete paragraphs.''}  Ordinarily, Miss Jones would ask students to share their essays alongside their ChatGPT transcripts to check compliance with her policy. Today, she decided to try \system. In the following, we walk through her analysis of three students' essays (Fig.~\ref{fig:all_essay}), 
Matilda, Lavender, and Bruce.

\subsection{Reviewing Matilda's Essay: Using AI for Transitions \& Transformation}
\begin{figure*}
    \centering
    \includegraphics[width=0.9\textwidth]{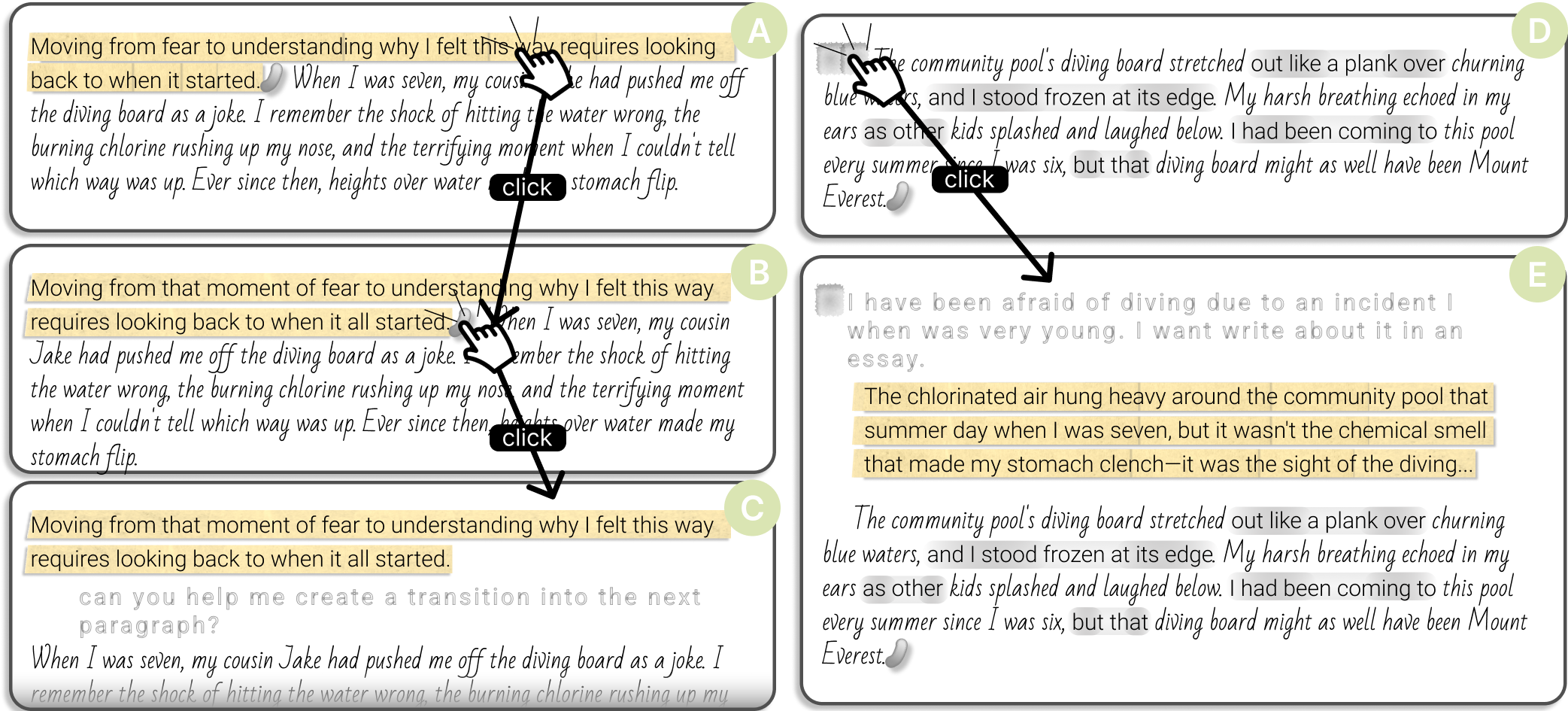}
    \caption{Usage scenario exploration of \protect\lightgreencircled{2A} Matilda' essay.}
    \label{fig:matilda}
\end{figure*}

When Miss Jones opens Matilda’s essay (Fig~\ref{fig:all_essay}A), she immediately notices several visual encodings (Fig~\ref{fig:matilda}A), signaling AI involvement at different points. Miss Jones prefers to read in two passes: first, to comprehend Matilda's writing, then to assess it against her rubric and prepare feedback. On her first pass, she notes smudge marks in the first paragraph and masking tape at the beginning of paragraphs two and three. Both represent AI-generated text, but with different functions: the masking tape indicates inserted content, while smudge marks show AI tone transformations of existing text. The masking tape metaphor suggests something provisional or externally added, and the 'crumpled' state further conveys whether the suggestions were kept intact, modified, or fragmented. This supports Miss Jones's rapid inference about the degree of human intervention. In her second pass, Miss Jones examines the crumpled masking tape in the second paragraph (Fig.~\ref{fig:matilda}B), indicating that Matilda deleted words within an AI insertion: \textit{``Moving from fear to understanding why I felt this way requires looking back to when it started.''} Clicking the tape reveals the original AI text for five seconds before it crumples again. Next, Miss Jones clicks on the eraser crumb beside this tape (Fig.~\ref{fig:matilda}C), revealing Matilda's ghost text prompt requesting help with a transition from the opening paragraph (Fig.~\ref{fig:matilda}D). She finds a similar transition request related to the masking tape in the third paragraph.  

At the start of the essay, multiple residual glue marks indicate attempts to generate full drafts with AI that Matilda later discarded. Clicking on one (Fig.~\ref{fig:matilda}E) shows the rejected AI text (Fig.~\ref{fig:matilda}F). Miss Jones infers that Matilda rejected the AI draft in favor of her own writing.  Finally, she examines the smudge marks in the first paragraph. Since these encode tone transformations of existing content, Miss Jones decides not to click into the underlying prompts.  From these traces, she concludes that: \textit{Matilda struggles with transitions, but is actively working to improve them, using AI as a scaffold rather than a crutch.} Her feedback encourages Matilda to experiment with drafting her own transitions before refining them with AI.  

\subsection{Reviewing Lavender's Essay: Using AI for Iterative Generation \& Feedback}
\begin{figure*}
    \centering
    \includegraphics[width=1\textwidth]{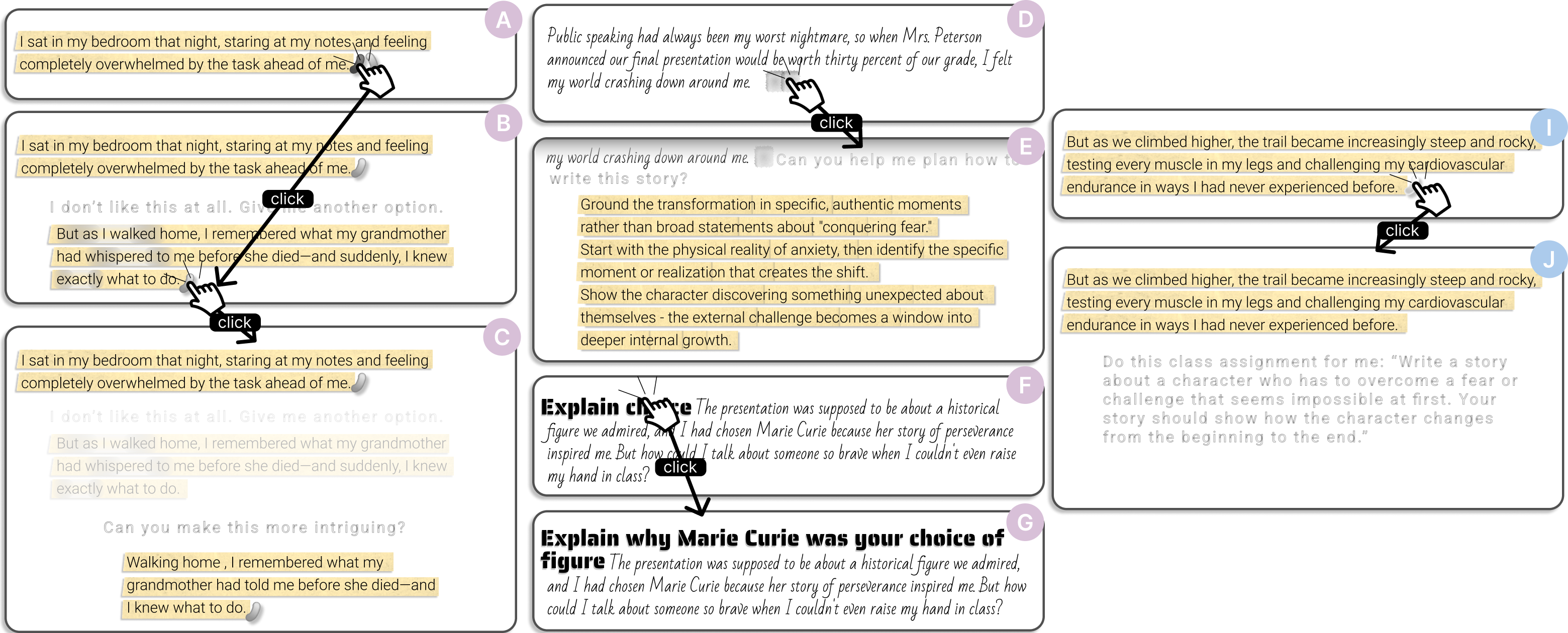}
    \caption{Usage scenario exploration of \protect\lightpurplecircled{2B} Lavendar and \protect\lightbluecircled{2C} Bruce's essay.}
    \label{fig:lav_bruce}
\end{figure*}

Lavender’s essay illustrates a different pattern (Fig~\ref{fig:all_essay}B). During her first pass, Miss Jones notices residual glue, masking tape, eraser crumbs, and stencil marks. Although few AI sentences remain in the final draft, Lavender appears to have leaned heavily on AI throughout the process. At the end of the first paragraph, a residual glue mark (Fig~\ref{fig:lav_bruce}D) reveals an AI-generated outline (Fig~\ref{fig:lav_bruce}E), ultimately deleted but clearly echoed in the final structure of Lavender's essay. The second paragraph (Fig~\ref{fig:lav_bruce}A) contains a masking tape with two eraser crumbs, one darker than the other. The darker crumb signals a more complex prompt. Clicking on it reveals one sentence in ghost text and another in masking tape (Fig~\ref{fig:lav_bruce}B). Because Lavender included a previously generated AI sentence in her prompt, this nesting creates a chain of iterations (Fig~\ref{fig:lav_bruce}A–C). Over the taped sentence \textit{``But as I walked home, I remembered what my grandmother had whispered to me before she died—and suddenly, I knew exactly what to do.''}, Miss Jones also sees segmented gray smudge marks, indicating recursive tone adjustments(Fig~\ref{fig:lav_bruce}B). Clicking on the associated crumb confirms that the AI was repeatedly asked to refine its own text (Fig~\ref{fig:lav_bruce}C).  

In addition, the third, fourth, and sixth paragraphs show stencil marks: solid in the third and fourth, hollow in the sixth (Fig~\ref{fig:all_essay}). These encode AI feedback requests, with solid marks showing integrated changes and the hollow mark showing rejected feedback. For example, in paragraph three, clicking the stencil mark (Fig~\ref{fig:lav_bruce}F) reveals the AI feedback that helped Lavender expand her reasoning for choosing Marie Curie (Fig~\ref{fig:lav_bruce}G).  From this move, Miss Jones concludes: \textit{Lavender is doubtful about her writing}, seeking reassurance through repeated generation and critique. This is less about laziness than about uncertainty: she is looking for confidence in her authorial voice. In her feedback, Miss Jones highlights moments where Lavender’s own drafts were strong without AI assistance, encouraging her to trust her instincts more than her writing process revealed that she did.

\subsection{Bruce: Using AI for Passive Writing}
Bruce’s essay shows yet another pattern (Fig~\ref{fig:all_essay}C). When Miss Jones opens it, she is struck by the overwhelming presence of masking tape. Two full paragraphs are entirely taped, which means that they were fully AI-written, while the third paragraph is only partially taped, with gaps showing Bruce’s own contributions. Throughout the essay, there is only a single eraser crumb at the end of the fourth paragraph. Its faint color suggests a simple, low-effort prompt. Clicking on it (Fig~\ref{fig:lav_bruce}I) reveals ghost text showing that Bruce had pasted the entire class assignment into ChatGPT (Fig~\ref{fig:lav_bruce}IJ, generating an essay with minimal input beyond one short addition in paragraph three.  

Unlike Matilda’s selective scaffolding or Lavender’s iterative exploration, Bruce’s work appears largely AI-written, with little evidence of his own thinking or revision. Although the essay looks polished, the skeuomorphic cues expose its shallow authorship. In her feedback, \textit{Miss Jones notes the lack of original writing and stresses the importance of Bruce developing his independent writing skills. She warns that overreliance on AI not only hinders his growth but also violates her classroom policy.}  

Taken together, these three cases illustrate how \system makes different modes of AI-assisted writing visible. Matilda uses AI selectively as a scaffold, Lavender iterates recursively to build confidence, and Bruce relies passively on AI for entire sections. By revealing these distinctions, \system allows teachers like Miss Jones to tailor the feedback to the needs of each student, something that would be impossible with plain text alone.  
\section{\systems Data Model}

\begin{figure*}
    \centering
    \includegraphics[width=\linewidth]{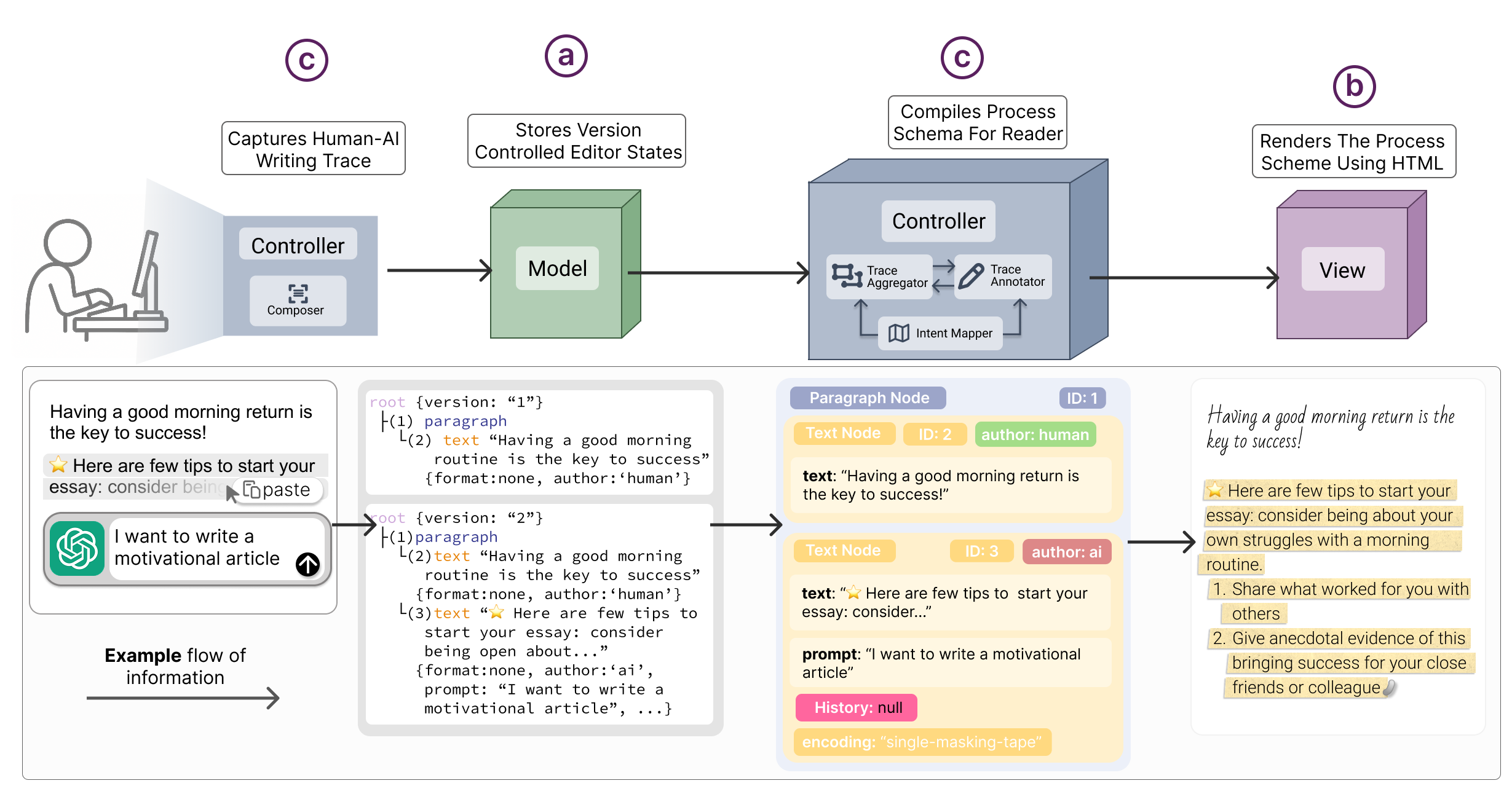}
    \caption{Overview of \system’s MVC architecture. (a) The model stores the writing process, (b) the view presents the process to readers, and (c) the controller manages the bi-directional logic between model and view.}
    \label{fig:system_architecture}
\end{figure*}

As shown in Figure~\ref{fig:system_architecture}, \system is implemented as a model-view-controller (MVC) architecture. Here we elaborate on the data model for \system (Figure~\ref{fig:system_architecture}a), and in (Sections~\ref{sec:view} and~\ref{sec:controller}) we discuss the other components. 

\begin{figure*}
    \centering
    \includegraphics[width=\linewidth]{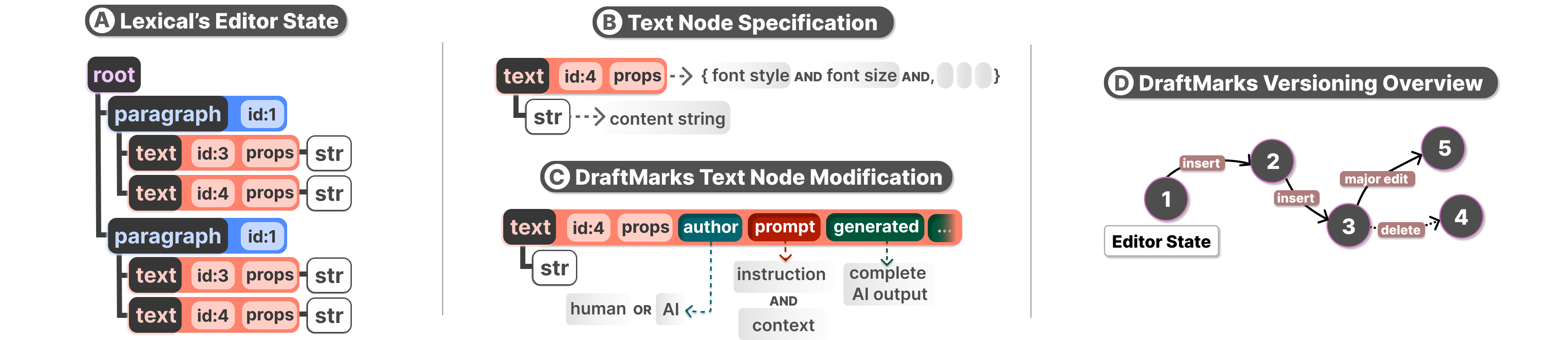}
    \caption{Here we show: (A) typical representation of editor states in rich text editors, (B) their corresponding text node, (C) text node modifications made for \system and (D) lastly overview of our version controlling architecture for capturing human-AI writing process trace. }
    \label{fig:editor_state}
\end{figure*}

Our system captures human-AI collaborative writing through version-controlled editor states that preserve the complete interaction history. In most modern rich text editors~\cite{prosemirror2025,lexical2025,tiptap2025,googledocs2025}, each document is stored as a hierarchical node-based structure, often called an editor state, as shown in figure~\ref{fig:editor_state} A. Nodes can be structural elements such as paragraphs, headings, lists, or text nodes that point to content strings (Figure~\ref{fig:editor_state} B). Structural elements such as paragraphs, headings, and lists contain sequences of text nodes, and each text node can include style properties (e.g., bold, italics). We extend this framework (Figure~\ref{fig:editor_state}C) by attaching provenance information directly to the text nodes: (1) \textit{Author}—whether the content was written by a human or generated by AI, (2) \textit{Prompt}—the instruction and context used when the text was AI-generated, and (3) \textit{Generated}—the complete AI-generated text. Often writers ask AI for feedback on their content without importing that feedback into their writing canvas; these AI-generations are stored in orphan AI-authored text nodes. 

\begin{figure*}
    \centering
    \includegraphics[width=\linewidth]{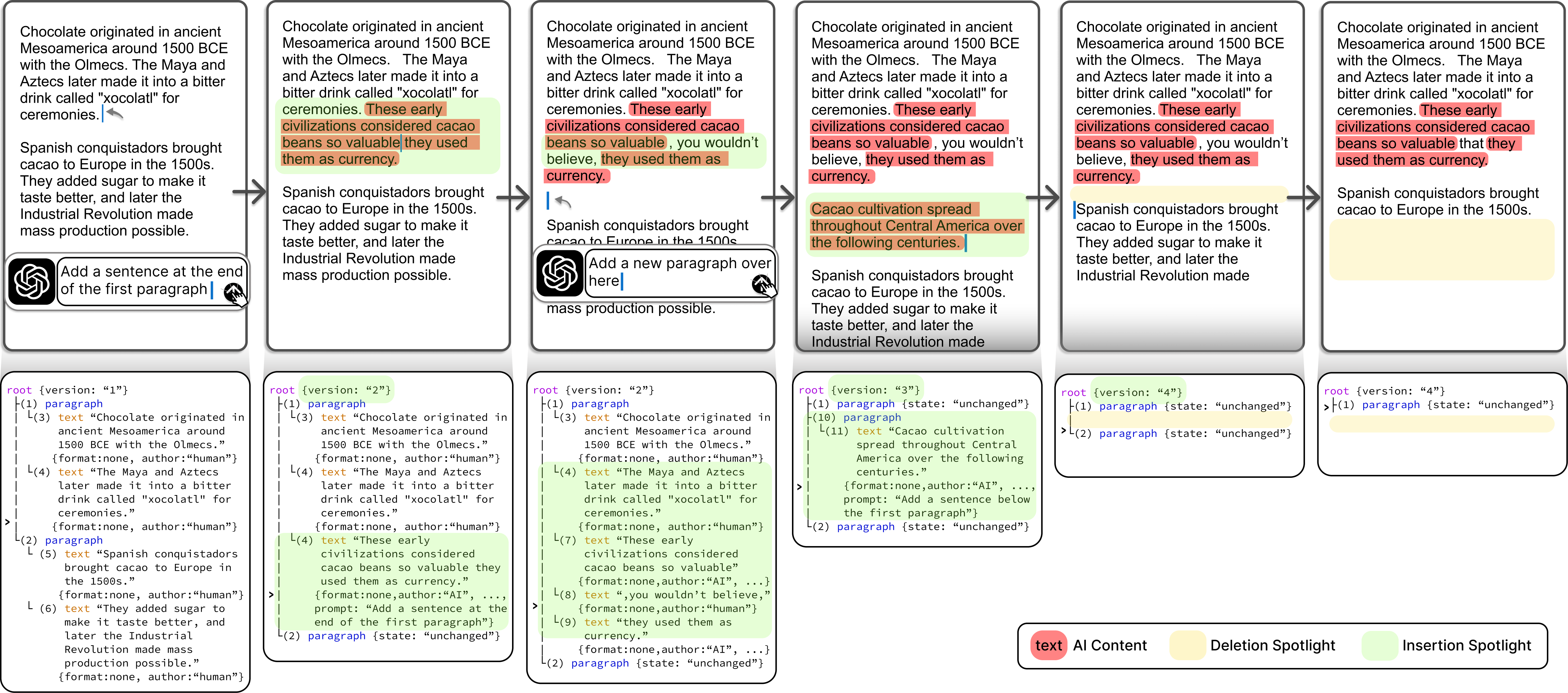}
    \caption{Here we demonstrate the changes occurring within the model of \systems on different points with human-AI writing collaboration. We ground this example in a writer using ChatGPT to write about the history of chocolate.}
    \label{fig:model_timeline}
\end{figure*}

Versioning in our system is event-driven rather than time-based. Whereas collaborative editors like Google Docs track synchronous and concurrent human-human edits through operation transformation~\cite{10.1145/289444.289469,googledocs2025} and periodic snapshots, our model reflects the sequential nature of the human-AI workflow (Figure~\ref{fig:editor_state}D). Current AI-assisted tools do not support concurrent human and AI editing~\cite{Siddiqui:2025:ScriptShift, Zhang:2023:VISAR}. A new version is created only when: (1) an AI-authored text node is \textit{inserted}, or (2) an AI-authored text node is \textit{fully removed}, or (3) an AI-authored text node has \textit{10+ characters deleted} from its content string (we use this threshold to capture micro edits). All other human revisions remain within the current version. 

\begin{figure*}
    \centering
    \includegraphics[width=\linewidth]{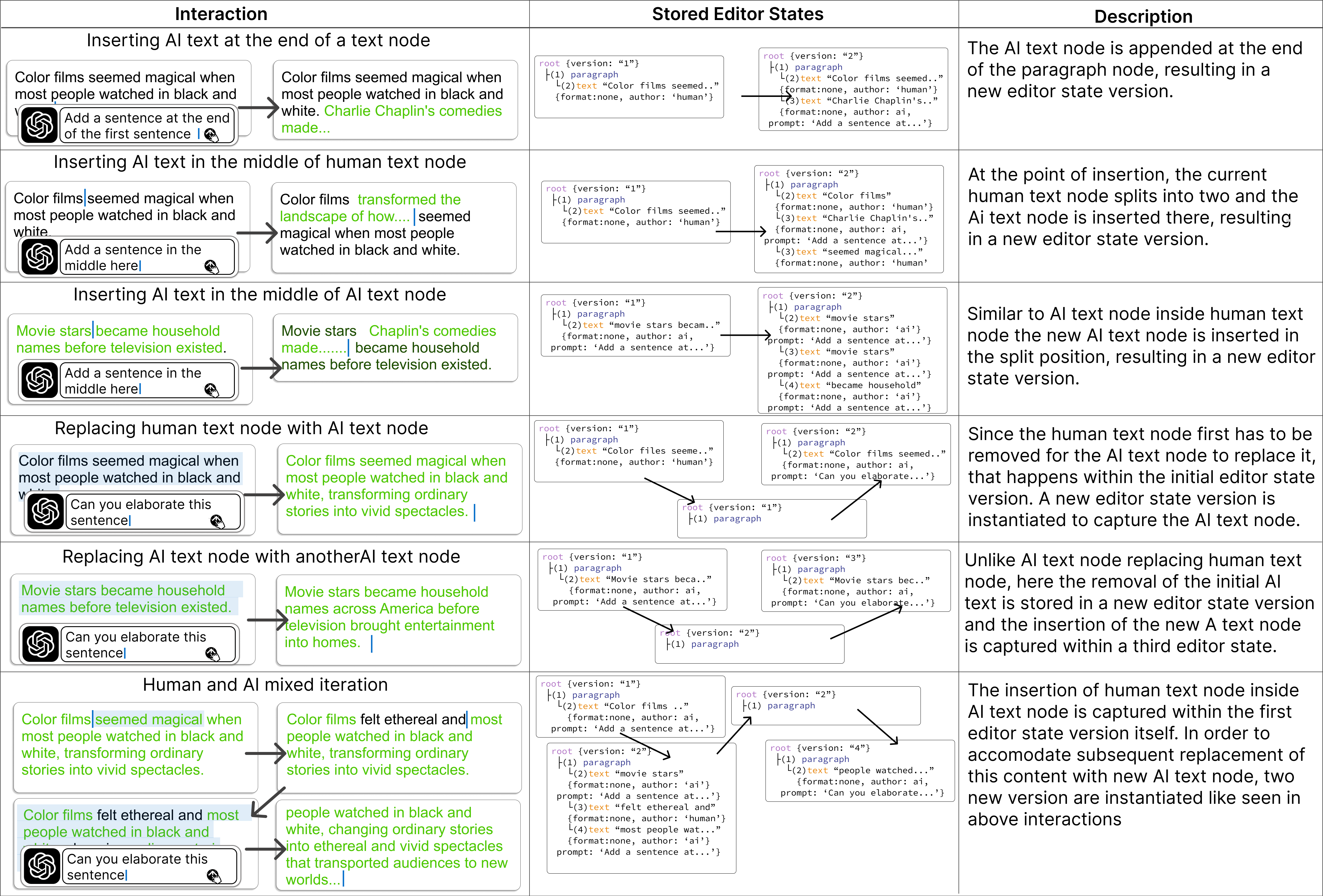}
    \caption{In this figure we show some common ways of integrating AI in writing. We use the example of ChatGPT being used in writing and their the corresponding overview of our version controlled data model. The text is green represent AI generation.}
    \label{fig:model_interaction}
\end{figure*}

An example of this flow is illustrated in Figure~\ref{fig:model_timeline}. From left to right, the content evolves through the collaboration between AI and human. The corresponding editor state is shown in the next row with green and yellow highlights indicating key changes such as node insertion and version change across editor states. To ensure efficient storage, each version is represented through \textit{references to} and not a \textit{copy of} the structural and text nodes, enabling compact history tracking similar to Git. This model design provides a fine-grained record of human–AI collaboration,  capturing both the provenance of AI contributions and the depth of the document revision. The details regarding the capture of this knowledge representation are presented in Sect.~\ref{sec:controller}. We also include the version-controlled editor representations for various types of human-AI interaction during writing in Fig.~\ref{fig:model_interaction}.

\begin{figure*}
    \centering
    \includegraphics[width=\textwidth]{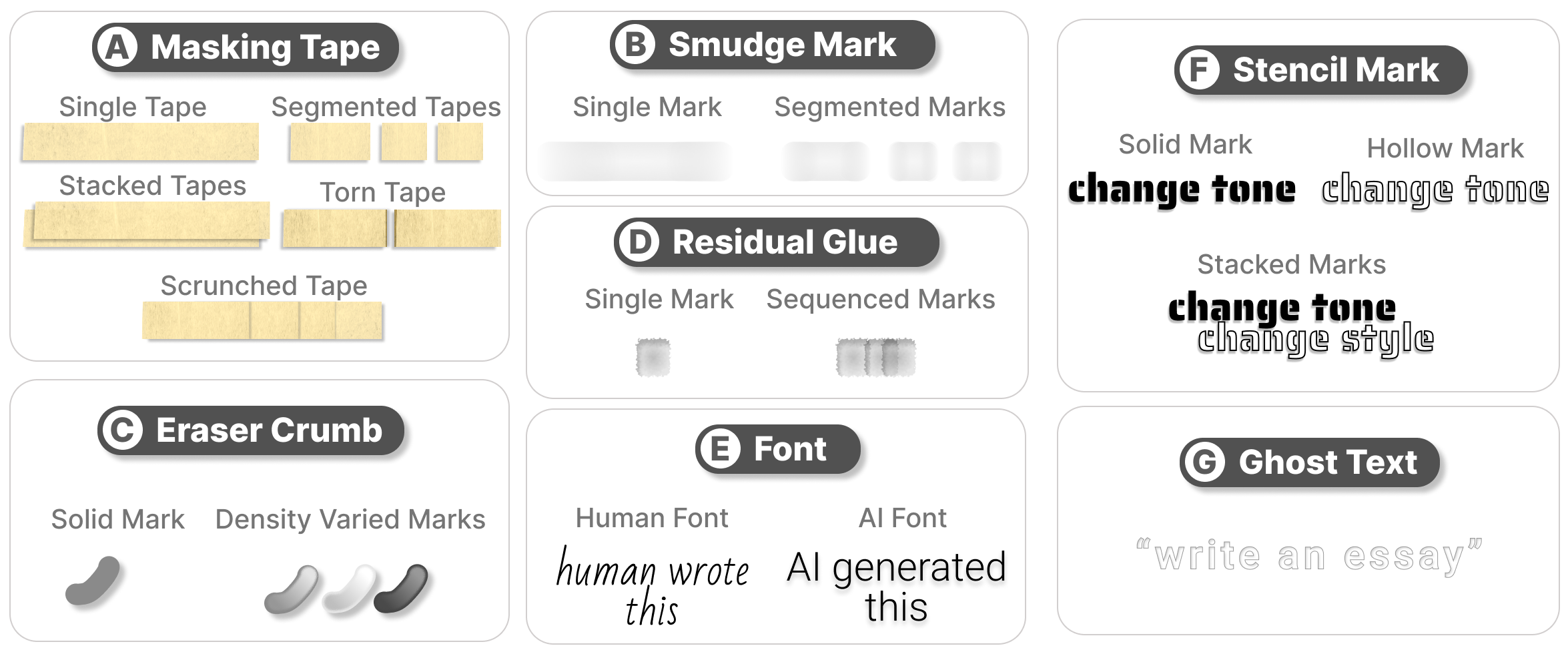}
    \caption{Marks in DraftMarks. Each addresses a unique challenge in visualizing temporal human-AI coediting.}
    \label{fig:marks}
\end{figure*}

\section{\systems View: Skeuomorphic Encodings} \label{sec:view}
In \system, we adopt skeuomorphism as a deliberate design choice to represent human-AI writing collaboration. By incorporating visual elements into the everyday physical experience, skeuomorphic designs can increase interpretability and reduce the learning curve for new systems~\cite{Lee:2022:BiosignalVisVR, Yao:2024:Taotie}. Our technique uses several metaphors to semantically map writing behavior to visual representations. The selection of visual channels and variants for representing process information is made by the controller module (Section~\ref{sec:controller}).

\textbf{Masking Tape.  } They (Fig.~\ref{fig:marks}A) indicate passages that were initially generated by AI. We selected this metaphor for AI-generated content since tape implies something provisional or temporary and externally added, qualities that align with how AI-generated content functions in co-authoring. Tape as a metaphor also allows crumpling or tearing of the tape which can visually communicate whether suggestions were kept intact, modified, or fragmented, supporting quick visual inference about the degree of human intervention. The tape also allows us to write over and add additional tape onto the new tape, all of which can signal iteration over the same content by repeated prompts. Tape specifically encodes new content generation, which was not part of the prompt passed to the AI. \system supports five different views for the masking tape: (1) \textit{single masking tape}: which represents a continuous strip of masking tape marking the content generated by AI, in this view granular details regarding exactly what ideas came from AI and which were already present in the prompt are not preserved as the masking tape strip marks the AI output without nuance; (2) \textit{stacked masking tape}: represent layered tapes for iterative AI generation, trying to make AI insertions at the same spot by replacing the prior AI suggestion over and over again; (3) \textit{scrunched masking tape}: which preserves information regarding word deletions made by humans within the AI text; (4) \textit{torn masking tape}: which preserves information regarding word insertions made by humans within the AI text; (5) \textit{segmented masking tape}: which preserves information regarding which phrases in the generation were AI original generation, and which ones were part of the prompt that the human wrote.

\textbf{Smudge Marks.  } They (Fig.~\ref{fig:marks}B) indicate passages that were modified by AI for tone shift, and not for the insertion of the original content. Smudges indicate areas of semantic drift, where the meaning of a sentence has changed substantially over revisions. We use smudging as a metaphor because it suggests motion, uncertainty, and transformation, qualities inherent to iterative meaning-making. \system can support two different views for the smudge mark: (1) \textit{single smudge mark}: which represents a continuous streak of smudge over the AI's output without disambiguating precise changes made by AI; (2) \textit{segmented smudge mark}: it preserves information regarding which phrases in the generation were AI original generation and which ones were part of the prompt that the human wrote.

\textbf{Eraser Crumbs.  } Eraser crumbs (Fig.~\ref{fig:marks}C) naturally evoke the act of erasing and rewriting by hand, activities closely related to the revision effort. In \system, the eraser crumbs appear next to AI smudge marks and masking tape to show the virtually erased prompts. The eraser crumb has two  views: (1) \textit{solid crumb} that has a uniform gray color; (2) \textit{ density varied crumb}: which has different shades of color. The differentiator between the two is nuance, in case of density varied crumbs the detail regarding complexity of prompts is encoded by the shade of gray.

\textbf{Residual Glue.  } Faint, translucent glue residue (Fig.~\ref{fig:marks}D) remains where masking tape was once--- to indicate that an AI suggestion was initially introduced--- but later removed entirely by the writer. Glue marks suggest absence with a trace, communicating that something was considered and discarded. Compared to showing the full deleted content, this subtle signal respects the integrity of the final draft while still surfacing decision points in the writing process. The views supported here are: (1) \textit{single glue mark}: showing only the last discarded AI generation at a passage; and (2) \textit{sequenced glue mark}: showing all versions of discarded AI generations at a passage.

\textbf{Font.  } While not skeuomorphic, \system uses contrasting font styles (Fig.~\ref{fig:marks}E) to signal the origin of the author. (1) Script or handwritten fonts for writer-authored text, and (2) sans-serif fonts for AI-generated content. Fonts provide an ambient but persistent signal that visually differentiates the origin of the text without resorting to color or other intrusive overlays.

\textbf{Stencil Marks.  } Light, hollow, or dashed letter forms are used to mark stenciled text (Fig.~\ref{fig:marks}F), passages where the writer incorporated AI feedback in a structured, guided manner. This metaphor draws from the educational practice of learning to write by tracing letters from a stencil or dotted line worksheet. This signifies that, while the structural scaffold came from the AI in the form of directed feedback, the final form was shaped by the writer. Stenciled text represents a middle ground between automation and authorship, highlighting human effort built on top of an AI scaffold. This encoding has 4 variants: (1) \textit{single stencil mark}: a single character mark at the margin of the paragraph; (2) \textit{layered stencil mark}: which marks multiple AI feedback generations over the same passage; (3) \textit{dotted font strokes}: the text in the passage next to the stencil mark appears dotted and not as solid lines to depict lack of AI feedback integration done by the writer; (4) \textit{lined font strokes}: this is the complement of dotted font strokes for cases where the writer integrates the AI feedback.

\textbf{Ghost Text.  }Translucent ghost lines of text (Fig.~\ref{fig:marks}G) show prompts or ideas that were generated, but that were not ultimately included in the final output. This metaphor suggests absence or residue - text that was once present in the writing process but was ultimately excluded or ``left behind.'' Ghosted text gives readers visibility into the writer's exploratory moves without disrupting the coherence of reading the final document. Here, the two variants are: (1) showing only instructions from the prompt; (2) showing the full prompt which includes instruction as well as context.

These cues provide a visual language for \system that aligns the semantics of the writing process with familiar visual metaphors from physical writing and editing. As seen in Figure~\ref{fig:lav_bruce}B---segmented smudge marks over single masking tape---- the encodings can be overlayed on top of each other to represent rich breadth of complex writing and editing workflows. Different combinations enable the human-AI writing process to be traced at different levels of depth, depending on the importance given to the final content, behind-the-scenes prompts, iteration, or precise phrase-level provenance. Each of these metaphors was selected through experimentation via encoding data and deliberation amongst the authors, both for its recognizability and its ability to encode the corresponding process.

\begin{table*}
    \centering
    \fontsize{8}{10}\selectfont
    \begin{tabular}{@{}ccccc@{}}
    \toprule
    \multicolumn{5}{c}{\cellcolor{gray!25}\textbf{(A) Teachers}} \\
    \midrule
    \textbf{ID} & \textbf{Experience} & \textbf{Discipline} & \textbf{Grade Level} & \textbf{AI Acceptance} \\
    \midrule
    P1 & 10+ years & History & Middle School & 3 \\
    P2 & 10+ years & English & Middle \& High School & 5 \\
    P3 & 7 years & English & High School & 5 \\
    P4 & 10+ years & English & Middle School & 3 \\
    P5 & 7 years & English & Elementary School &  4 \\
    P6 & 9 years & History & Middle \& High School & 5 \\
    P7 & 10+ years & English \& Math & Middle School  & 4 \\
    \midrule
    \multicolumn{5}{c}{\cellcolor{gray!25}\textbf{(B) Academic Reviewers}} \\
    \midrule
    \textbf{ID} & \textbf{Experience} & \textbf{PhD Year} & \textbf{Discipline}  & \textbf{AI Acceptance} \\
    \midrule
    R1 & 3  papers & 3rd Year & HCI& 4 \\
    R2 & 10+  papers & 6th Year & Ubicomp  & 3 \\
    R3 & 10+ papers  & 3rd Year & Robotics & 2 \\
    R4 & 6  papers & 3rd Year & Machine Learning & 4 \\
    R5 & 3  papers & 3rd Year & Machine Learning & 3 \\
    R6 & 7  papers & 5th Year & HCI & 4 \\
    R7 & 6  papers & 4th Year & HCI & 5 \\
    \midrule
    \multicolumn{5}{c}{\cellcolor{gray!25}\textbf{(C) General Readers}} \\
    \midrule
    \textbf{ID} & \textbf{Reading} & \textbf{Education} & \textbf{Profession} & \textbf{AI Acceptance} \\
    \midrule
    S1 & Very often & High School Diploma & Undergraduate Student & 4 \\
    S2 & Very often & High School Diploma & Undergraduate Student & 4 \\
    S3 & Often & High School Diploma & Undergraduate Student & 5 \\
    S4 & Somewhat often & Bachelor's Degree & Engineer & 4 \\
    S5 & Very often & High School Diploma & Undergraduate Student & 5 \\
    S6 & Very often & Master's Degree & PhD Student & 5 \\
    S7 & Occasionally & High School Diploma & Undergraduate Student & 5 \\
    \bottomrule
    \end{tabular}
    \vspace{1em}
    \caption{Participant demographics from our controller design probe (Sect.~\ref{sec:controller_model_to_view}). ``AI Acceptance'' refers to responses on whether AI use in writing is acceptable, rated on a 5-point scale (1 = strongly disagree, 5 = strongly agree).}
    \label{tab:formative_demographics}
\end{table*}

\section{Controller Logic: Generating Skeuomorphic Encodings from Data Model}
\label{sec:controller}
Reading practices are fundamentally \textit{contextual}. Different stakeholders have different interpretive frameworks and evaluative priorities. Teachers engage in pedagogically-oriented reading --- simultaneously assessing content comprehension and student learning processes --- while academic reviewers focus on novelty, methodological rigor, and scholarly contribution. These stakeholder-specific requirements should be taken into account when visualizing human-AI writing collaboration. In other words, it is difficult to formulate a singular augmentation algorithm that can completely capture all insights different readers might need while also remaining legible to the reader. Therefore, in \system we take a human-centered approach to design and develop the various algorithms for mapping the data model to role-dependent views. These algorithms are implemented in \systems controller module (Figure~\ref{fig:system_architecture}c).

\subsection{Formative Study for Designing Controller Logic}

To better understand reader needs, we conducted a formative study with three different stakeholders: (A) teachers, (B) academic reviewers, and (C) general readers. From this diverse pool of readers ($N=21$, i.e., 3 stakeholder x 7 participants), we sought to understand what visibility into the writing process they might need, and the associated comprehension and assessment-related decisions they would make. To conduct the study, we developed a baseline version of \system implementing the different skeuomorphic augmentations mentioned in Section~\ref{sec:view}. This version of the system is intended as a design probe for the study. \system implemented \textit{masking tape} to represent AI content,  \textit{ghost text} to represent prompts, and additionally \textit{eraser crumb}, \textit{smudge marks}, and \textit{residual glue}, along with different fonts for human-written and AI-generated content. This baseline tool reflects a deliberately simple implementation, guided by heuristics and our design judgment, to surface possibilities rather than optimize performance.

\medskip
\noindent\textbf{Participants.  }
We recruited participants via LinkedIn, Twitter, and word of mouth. For \textit{teachers}, we had the selection criteria of being a teacher currently and having long-form writing (narrative or argumentative) assessments as part of their curriculum. All teachers recruited were from private schools with relatively small class sizes (20 classrooms of students). For \textit{ reviewers}, we had the selection criteria of being a senior Ph.D. student who has reviewed at least three conference papers. For \textit{general reader}, we recruited participants who did not meet the inclusion criteria of the other two stakeholders. All details regarding the participants can be found in Table~\ref{tab:formative_demographics}. Each participant was compensated with \$30 for their time. The institution's IRB approved the study. 

\medskip
\noindent\textbf{Study procedure.  }
Each study session lasted no more than one hour. At the start of each session, participants received a live demonstration of \system on an example text that showcased all the basic skeuomorphic encodings and their associated interactions. After this introduction, participants were asked to assess a piece of writing ---co-written with AI--- relevant to their readership. The details regarding essay selection are provided in the following paragraph. The participants were asked to think aloud about their assessment and comprehension. After this activity, participants received a general rubric to assess the writing on various dimensions. The purpose of the rubric scoring was only to understand their evaluative decision-making processes. Participants were then shown the video recording of the actual writing process from the authors of the essays to elicit any additional information they would have liked to see in \system to help their comprehension and assessment needs. Lastly, we administered a demographic survey and a short questionnaire specific to each readership.

\medskip
\noindent\textbf{Essay Selection. }
For the study task, we curated role-specific essays. For teachers, we selected a writing session from the CoAuthor dataset \cite{lee2022coauthor}, which captures interactions with a tool that generates four alternative sentences as continuations to the current text based on writer-initiated triggers. For reviewers, we recruited a postdoctoral fellow who wrote an article aimed at a trade magazine audience—in this case, ACM Interactions—based on their dissertation. The participant used Script\&Shift~\cite{Siddiqui:2025:ScriptShift}, an AI co-writing tool that employs a collage-based metaphor with multiple ``AI-helpers'' specialized for different writing subprocesses. After a demonstration of Script\&Shift, the participant completed the task and received a \$30 honorarium. For general readers, we recruited a master’s student to write an article on ``privacy in the digital age'' with the assistance of ChatGPT. In the case of CoAuthor, we obtained the writing session video from the project’s website. For Script\&Shift and ChatGPT, we recorded participants’ screens while they wrote. Additionally, for the recruited writers, we ran the \system composer in the background of their browser to capture the data model.

\begin{figure*}
    \centering
    \includegraphics[width=0.9\textwidth]{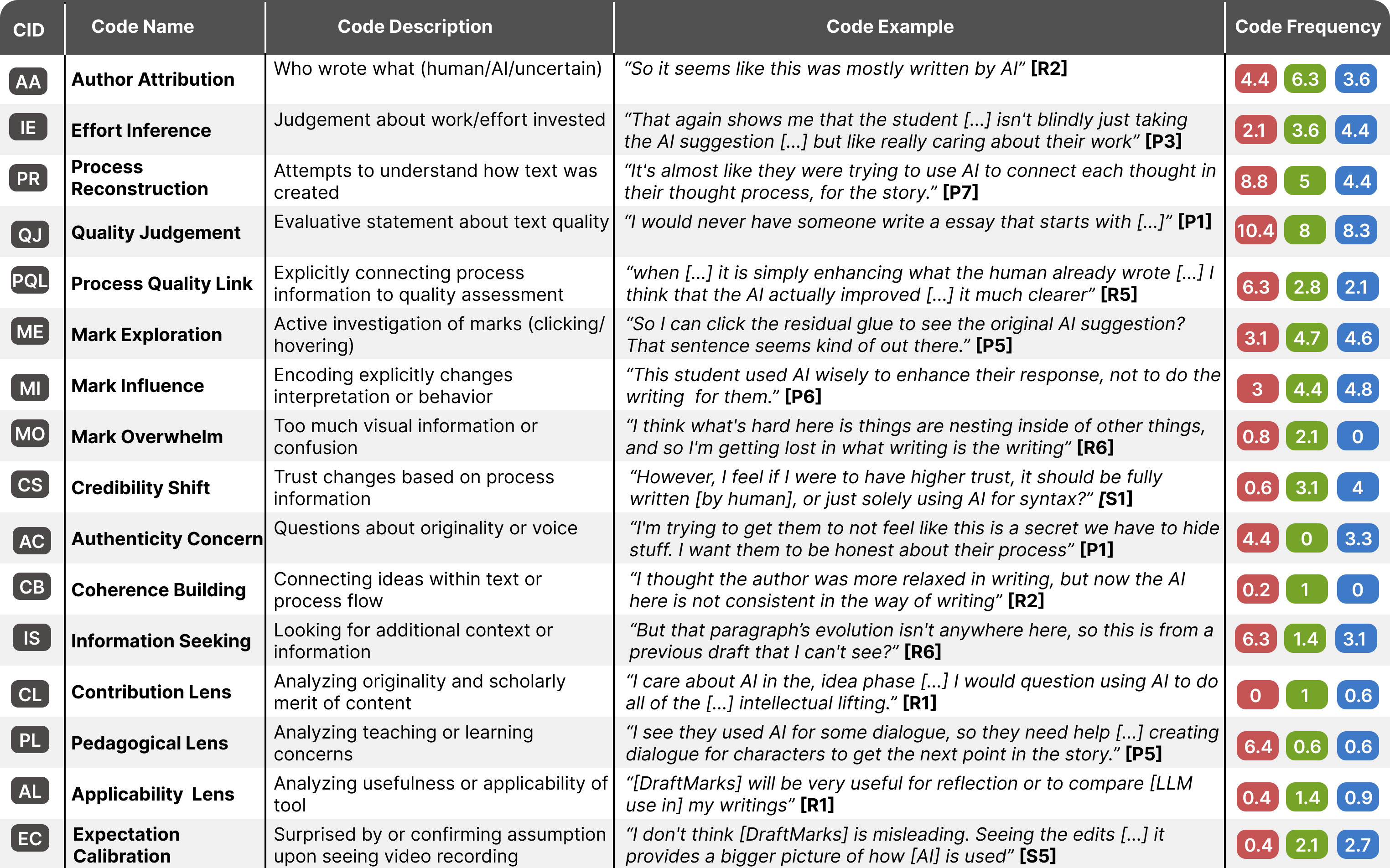}
    \caption{This figure showcases the final code book devised from thematic analysis of all 21 transcripts across stakeholders. The code frequency column contains average frequency of the code across participants for teachers (red), reviewers (green),and general readers (blue).}
    \label{fig:codebook}
\end{figure*}

\medskip
\noindent\textbf{Data Analysis.  }
We conducted reflexive thematic analysis~\cite{BraunClarke2006}. Two of the authors collaboratively coded six transcripts and video recordings to develop an initial codebook, which was then applied to the remaining transcripts. At this phase, the coding was done independently by the authors using the initial code book. The authors were invited to come up with new codes that were not observed in the initial subset. The final codebook, devised by the authors, was combined into a single, cohesive one after a round of discussion (number of top-level codes = 19). An independent rater not affiliated with the article then assisted in the process of combining, splitting, or removing existing codes (number of top-level codes = 16). The final code book constituted the evaluative and interpretative process traces we were looking for, as shown in Figure~\ref{fig:codebook}. 

Following this activity, the two authors and the independent rater mapped these codes to the different variants across the visual encoding channels until agreement was reached. Three codes: (QJ) quality judgment, (AL) applicability lens, and (EC) expectation calibration were excluded from this process. QJ was excluded because it was directly related to the task and, therefore, was evenly present in high frequency for all participants. We exclude AL as this was the participants who commented on how \system could be helpful for different applications, not necessarily offering any visual encoding or specific insight into human-AI collaboration. Lastly, we exclude EC because of their reaction to seeing the video. All remarks about \system and their reading needs during the video showcase were captured in different codes.

\begin{figure*}
    \centering
    \includegraphics[width=0.7\textwidth]{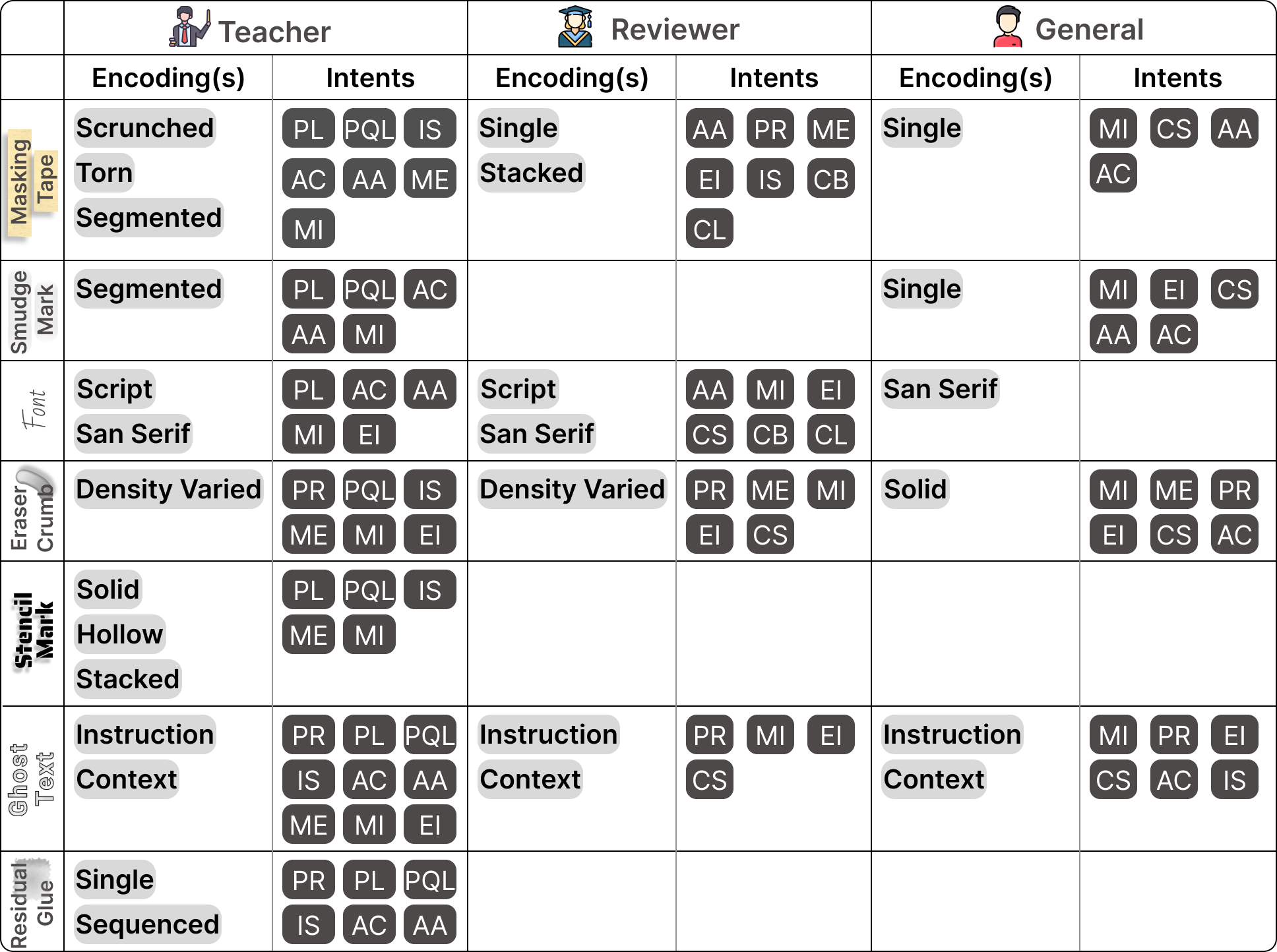}
    \caption{This figure encapsulates findings from the formative assessment. For each stakeholder, the visual encoding variants available to render are listed along with the corresponding reading intent they satisfy. This logic drives the intent mapper in the Model-to-View Controller}
    \label{fig:controller_table}
\end{figure*}

\subsubsection{Study Results}
Overall, we observed that teachers were more curious about the trace behind the written content than the other two stakeholders. Reviewers shared that they often have limited time to review papers, and the trade-off between cognitive load and insights was not justifiable for them in most instances. They did say that provenance of ideas is helpful and expressed desire for more summative insights in contrast to formative pieces in \system. Lastly, for general readers \system had an impact on their trust and willingness to read to article, sharing they value authenticity and originality in author's voice and \systems traces help capture that.

\medskip
\noindent\textbf{Insights on Teachers.}
We observed significant frustration among teachers regarding the monitoring and guidance of student AI use in assessments. P6 described existing detection mechanisms as unhelpful, noting that students use so-called ``humanizers'' to circumvent tools like GPTZero. However, teachers demonstrated pragmatic acceptance of AI as an educational reality. P6 shared, \textit{``I came to the conclusion that [AI] is here to stay, that it is powerful. I am using it! [...] If I open the conversation up and let them show me how they're using it, I can start to figure out how to help them.''} Teachers emphasized that AI should support rather than replace student thinking. P4 explained, \textit{``The point of writing is to express your thinking—to make visible the invisible thinking. So AI should help you make your thinking visible. It shouldn't do your thinking for you.''} With respect to \system, teachers valued its formative assessment potential. P2 shared that the tool\textit{ ``makes it clear what the student is struggling with from a formative point of view. That to me feels like a point that I like as a teacher.''} However, teachers noted that such detailed process analysis requires manageable class sizes and sufficient grading time. Teachers' emphasis on formative assessment and process reconstruction (PR in ) directly informed our controller design. Since teachers frequently engaged in pedagogical lens evaluation (PDL in Figure~\ref{fig:codebook}) and connected AI traces to quality judgments (PQL in Figure~\ref{fig:codebook}), we implemented multiple encoding variants for each channel to support granular process analysis (Figure~\ref{fig:controller_table}). For example, teachers' interest in students' decision-making around rejected AI suggestions (residual glue) led us to preserve these traces for even when content was discarded. The controller's aggregation depth for teachers is set to capture fine-grained interactions because, as P2 noted, understanding ``what the student is struggling with'' requires visibility into micro-level writing decisions. The teachers shared they had been experimenting with revision surfacing tools~\cite{brisk_inspect_writing,revision_history} for that micro-level decision making visibility. 

\medskip
\noindent\textbf{Insights on Reviewers.}
Academic reviewers expressed considerably less concern about AI usage in manuscript preparation, consistently emphasizing that evaluation should focus on the final writing product rather than the writing process. R7 articulated this stance clearly: \textit{``I'm here to review a piece of writing. [...] this seems like extra work.''} Reviewers demonstrated pragmatic acceptance of AI usage, with R7 stating they weren't bothered by generative AI in research papers. However, they distinguished between different types of assistance, explaining: \textit{``I don't care about AI in the writing, I care about AI in the idea phase.''} This reflects reviewers' concern with intellectual contribution over mechanical writing assistance. Despite general resistance to process-focused tools, some reviewers recognized potential value in specific circumstances. R6 noted it \textit{``would be helpful to have differentiation between when AI produces new ideas and when it is just helping polish content,''} particularly for non-native English speakers using AI for grammatical corrections. Reviewers' focus on intellectual contribution over process led to a minimal controller implementation. Their primary concern with author attribution of ideas (AA) drove our decision to highlight only content generation that might represent new intellectual contributions versus mechanical assistance. The sparse mapping in Figure~\ref{fig:controller_table} reflects reviewers' explicit preference against process exploration (ME in Figure~\ref{fig:codebook})--- we preserve only traces that support their core review obligations. For instance, iteration depth markers help reviewers quickly identify content the author invested significant effort in (PR), while other process details are suppressed to avoid the ``extra work'' R7 described. The controller prioritizes efficiency, surfacing only process cues that influence reviewer judgment (MI) without requiring deep engagement with collaborative traces.

Considering the overall mild interest in process trace of human-ai co-writing, our mapping of the reviewer's intents to visual encoding variants is quite sparse (Figure~\ref{fig:controller_table}). For support their review process, they are concerned with author attribution of idea (AA in Figure~\ref{fig:codebook}). They care about process reconstruction (PR) to see if there phrases/sentence the author paid special interest to (high iteration depth). They seemed content relying on process cues to influence their judgment (MI in Figure~\ref{fig:codebook}) but largely believed any further process exploration through marks would be unhelpful (ME in Figure~\ref{fig:codebook}) as it takes away from their reviewing time.

\medskip
\noindent\textbf{Insights on General readers.}
For general readers, AI acceptance was contingent on transparency and authorial effort. Unlike professionals with specific obligations, readers expressed concerns about authenticity and personal investment. S6 articulated this sentiment: \textit{``I don't mind AI tools [...] But it should be used as supplement to your writing, and not... It shouldn't do the writing for you.''} Readers valued human voice and perspective, with S6 explaining they read articles \textit{``to listen to people''} but extensive AI generation made them feel they weren't \textit{``listening to someone's point of view''} but rather \textit{``what a chatbot wrote back to them.''} The perception of effort significantly influenced reader trust and engagement, with low-effort AI use raising questions about whether the content was worth reading. Regarding \system, readers found the visualization valuable for understanding the writing process and desired comprehensive information about AI interactions, including access to original prompts and AI responses. General readers' authenticity concerns and trust responses were key in designing their controller. The dramatic trust reduction we observed when readers saw masking tape (as S6 demonstrated) led us to provide both smudge marks and masking tape encodings to them--- distinguishing between tonal refinement and novel content generation (Figure~\ref{fig:controller_table}). The aggregator maintains higher temporal depth for general readers than reviewers, reflecting their willingness to explore process details when transparency supports their authenticity assessments.

\medskip
\noindent\textbf{Writer Perspective.  }
Both reviewers and readers saw \system as a powerful metacognitive tool for writers themselves. Since writers often struggle to gauge AI effectiveness, process visualization becomes crucial for improving future collaboration. R1 explained: \textit{``Well, it might be very useful for the authors [...] I can even do the reflection and compare my writings to what the LLMs added [afterward].''} The tool also addresses temporal challenges in writing workflows. S5 observed: \textit{``if I write this, and then I come back, like, a week or two weeks later, I'm not sure I'll remember what I said [in prompts], right?''} However, stakeholders emphasized that writers should maintain control over captured content, noting privacy concerns about exposing all writing processes.

\subsection{Controller Implementation}
The controller facilitates bi-directional communication between the model and the view as shown in Figure~\ref{fig:system_architecture}. The controller is responsible for performing mutations to the underlying data structure of \system based on the input of the user (Section~\ref{sec:controller_view_to_model}). This is the part of the system where our version-controlled editor states are manipulated, corresponding to human-AI collaborative writing activity. The controller is also responsible for translating this data structure into a format ready to be rendered by the view and interacted with by the user (Section~\ref{sec:controller_model_to_view}).

\subsubsection{View to Model}
\label{sec:controller_view_to_model}
The capture of the data structure is handled by the composer module. The composer in \system is an extension of Lexical's editor state listener. The critical implementation artifact from \system is the inclusion of a versioned history manager and content provenance (human or AI) tracker. This module can store writing trace between AI and human for several common AI co-writing setups: (1) split context (Google Doc with ChatGPT), (2) integrated co-writing tools (Script\&Shift, ABScribe~\cite{reza2024abscribe}), and (3) ambient AI assistants (Grammarly chrome extension). In line with findings around writer privacy in our controller formative study, the composer requires explicit user permission for writing tracking, preventing tracking of content they are not comfortable sharing. The composer tracks keystroke-level data; content it can map to keystrokes is marked as human content, whereas that which it cannot is considered AI-generated. Copy and pasted content that is not local to an app is also considered AI-generated. Through these heuristics, the composer acquires the ground-truth data model, which the model-to-view controller then processes to determine final visual artifacts.

\subsubsection{Model to View}
\label{sec:controller_model_to_view}
Based on the findings from our formative study, we devised context-specific algorithms for converting the data model into \textit{view} ready representations. Using our codebook, we distilled a set of priorities for each stakeholder and based on them identified mapping to encoding variants within the various visual encoding channels of \system. Our mapping between encoding channels and stakeholder is shown in Figure~\ref{fig:controller_table}.  

\begin{figure*}
    \centering
    \includegraphics[width=1\textwidth]{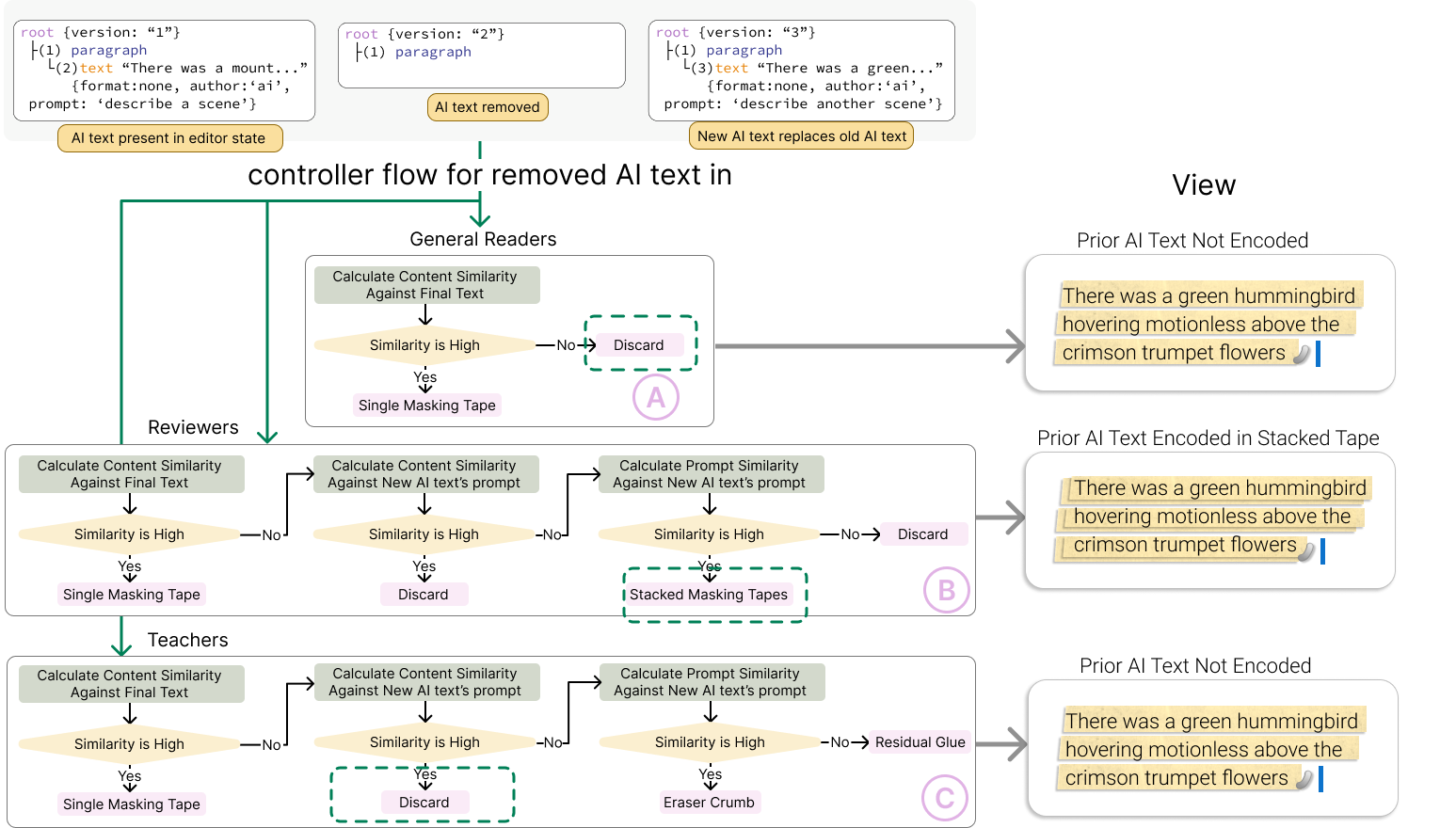}
    \caption{In this figure we show the controller's decision making trace from the model to the view. We illustrate how a previously discarded text node is aggregated and annotated for rendering based on the readership. The dashed green rectangle represented the rendering decision made for each reader type.}
    \label{fig:controller_example}
\end{figure*}

The controller transforms version-controlled editor states into a Process Schema—a monolithic nested structure that encapsulates all relevant traces for the reading stakeholder. This transformation is orchestrated by three interconnected components: the Trace Aggregator, Trace Annotator, and Intent Mapper, which work together to create a stakeholder-specific representation optimized for visualization. An example flow of decision making within the controller for different reader types is shown in Figure~\ref{fig:controller_example}

\begin{figure*}
    \centering
    \includegraphics[width=1\textwidth]{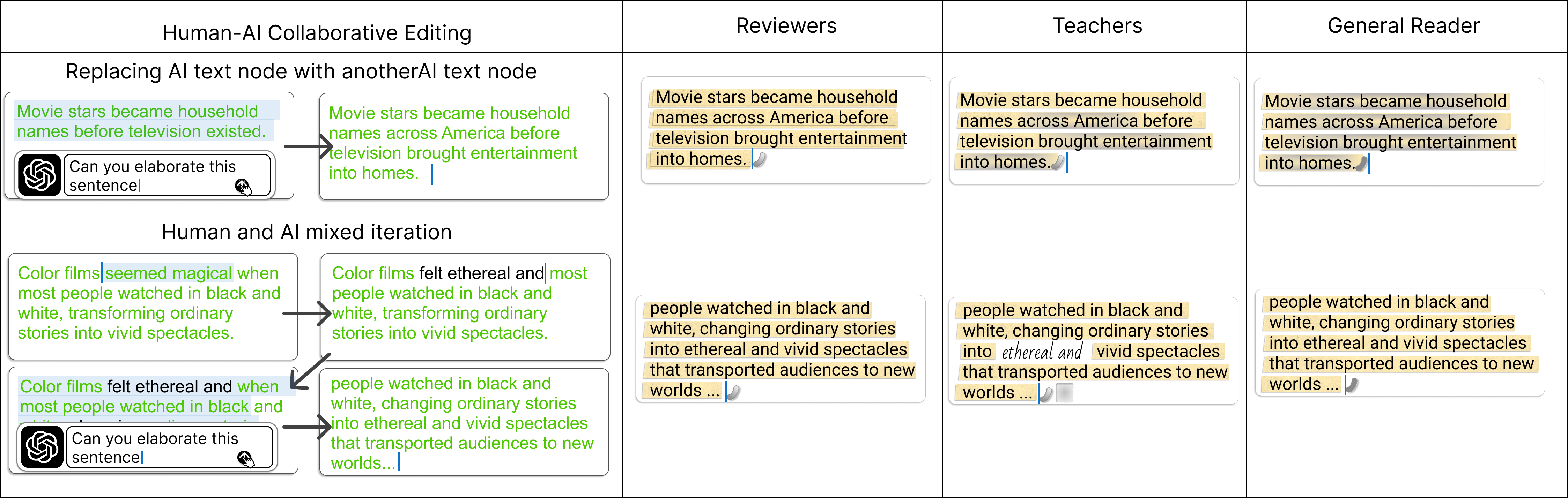}
    \caption{In this figure we show the more complex interactions from Figure~\ref{fig:model_interaction} (bottom two) and show what the view for each stakeholder would look like determined by the controller.}
    \label{fig:controller_view}
\end{figure*}

\medskip
\noindent\textbf{Intent Mapper.  }
The \textit{Intent Mapper} serves as the configuration hub, defining transformation parameters based on stakeholder profiles identified in our formative study. Teachers need granular access to student-AI interactions to assess learning and provide feedback, while general readers prefer streamlined views that highlight key collaborative patterns without overwhelming detail. The Intent Mapper maintains mappings between stakeholder intents and available visual encoding channels and their variants, communicating depth requirements to the Trace Aggregator and encoding permissions to the Trace Annotator. Details regarding this are present in Figure~\ref{fig:controller_table}. As shown in row 1 of Figure~\ref{fig:controller_view}, in case of reviewers the AI iteration is represented with stacked masking tape. The same iteration is rendered differently for teachers and general reader where further subdivision between tonal-shift and new-content generation type AI collaboration can be differentiated. In case of row 2 in Figure~\ref{fig:controller_view}, while reviewers and general reader have the same encodings, teacher is able to capture the discarded suggestions using residual glue.

\medskip
\noindent\textbf{Trace Aggregator.  }
The \textit{Trace Aggregator} determines which details from the version-controlled editor states should be preserved in the Process Schema (see Figure~\ref{fig:controller_example}). Rather than storing complete version histories, it makes selective decisions about temporal depth (how many versions back to preserve traces), granularity level (which text nodes warrant individual tracking), and nesting structure (how to organize traces hierarchically). The significance of changes and stakeholder type drive these decisions. At each inclusion/exclusion decision point, the aggregator consults the \textit{Intent Mapper} to ensure the preserved detail level aligns with the stakeholder requirements, then invokes the Trace Annotator (yellow diamond shape in Figure~\ref{fig:controller_example}) to determine the appropriate labeling.

\medskip
\noindent\textbf{Trace Annotator. }
The \textit{Trace Annotator} analyzes AI-authored text nodes to classify the nature of human-AI interactions and assigns appropriate visual encodings. It distinguishes between iterative and novel generation requests by examining prompt content and AI-generated text. Iterative AI calls that involve sequential requests for similar content can be represented as stacked masking tapes. Novel requests involving different content types receive separate visual elements. The annotator also identifies different generation types: new content generation is encoded with masking tape, tonal refinement uses smudge encoding, removed AI content appears as residual glue, and used prompts become clickable eraser crumbs that reveal ghost text. The annotator also passes to the Aggregator details about the consolidated structure of some content. Throughout this process, the annotator consults the Intent Mapper to ensure selected encodings align with the stakeholder's comprehension needs.

\begin{figure*}
    \centering
    \includegraphics[width=0.8\linewidth]{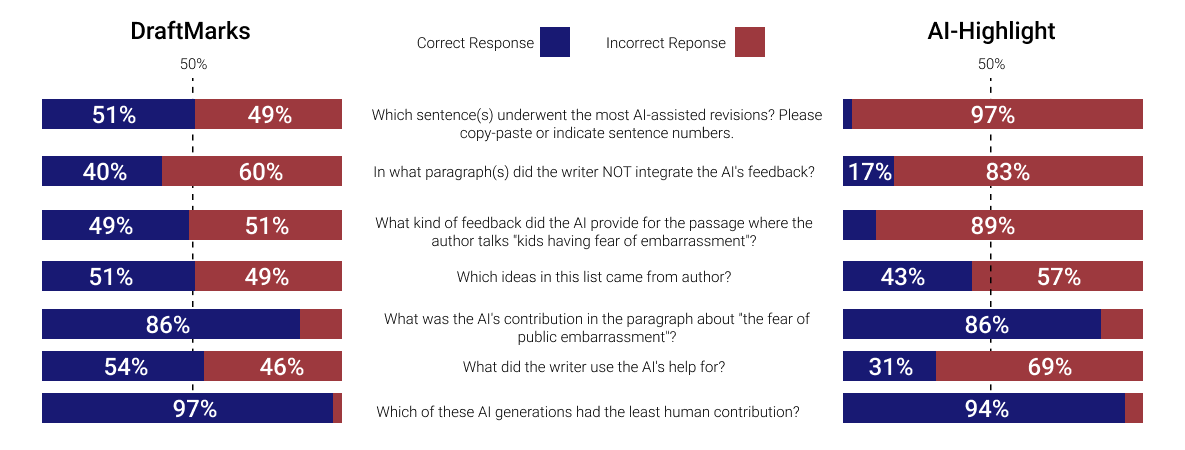}
    \caption{This figure showcases the distribution of correctness for the comprehension questionnaire between AI-Highlighted and \system condition}
    \label{fig:score}
\end{figure*}

\begin{figure*}
    \centering
    \includegraphics[width=0.8\linewidth]{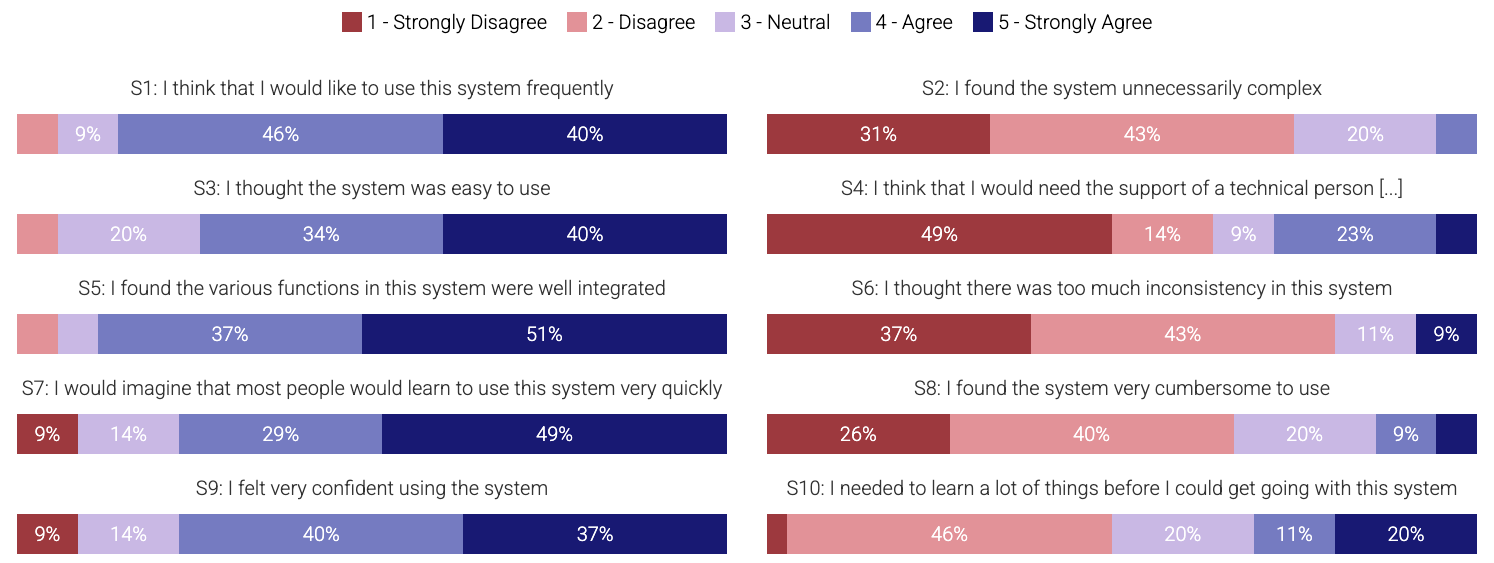}
    \caption{This figure showcases the distribution of SUS ratings (1-5) across the 35 participants for \system. For items in the left column, higher is better, and for items in the right column, lower is better.}
    \label{fig:sus}
\end{figure*}

\section{Evaluation}
Our study sought to evaluate the effectiveness of \systems visual encodings in surfacing human-AI writing collaboration patterns and enhancing users' comprehension and assessment abilities when reviewing AI-assisted content. We used a between-subjects design to compare our augmented reading tool against a baseline interface (described in Section~\ref{sec:study_procedure})

\subsection{Participants} 
We conducted the study through Prolific and screened to select participants who regularly engage in written content professionally, with roles including teacher, journalist, and copywriter. We also required participants to have a 95\% or higher approval rating on Prolific to ensure high-quality participation. We recruited 70 participants (F=36, M=32, NB=2), with 35 participants randomly assigned to each condition. Participants reported ages between 18 and 60 years and older, with the largest group (20 participants) in the age range of 31-40. Regarding AI usage in their own writing, most participants reported using AI sometimes (25) or often/very often (31), with only 2 participants reporting never using AI tools.

\subsection{Study Procedure} 
\label{sec:study_procedure}
Each study session lasted approximately 30 minutes and consisted of three phases: interface tutorial, practice task, and main evaluation task. The main task required participants to comprehend a 600-word essay about ``Need for Social Media Reform for Children's Use'' that was written collaboratively with ChatGPT and exhibited extensive human-AI interaction patterns. The participants then answered comprehension questions related to the content of the essay and the collaboration patterns. The control condition used a split view interface that shows the essay with AI-generated sentences highlighted alongside the available chat log. The treatment condition used \system with enhanced visual encodings for human-AI collaboration patterns. Following the main task, the treatment participants completed the Cognitive load survey and the system usability scale questionnaire. Participants in both conditions completed the demographic survey, filled out 3 transparency questions on a 7-point likert scale: (Q1) \textit{``I feel confident in my assessment of the essay''}, (Q2) \textit{``I felt I had enough information to assess the author's writing process''} and (Q3) \textit{``I was able to perceive the human effort in this essay''}  and provided open-ended feedback to these questions: (F1) \textit{``What, if anything, helped you most in understanding the human–AI collaboration?''}, and (F2) \textit{``What, if anything, was confusing or misleading?''}. Each participant received \$6 compensation for their participation.

\subsection{Results}
\subsubsection{Comprehension Questions.}  We analyzed question response accuracy using Fisher's Exact Test for individual questions (due to below 5 frequencies for correctness/incorrectness in some questions) and the Mann-Whitney U test for aggregate performance across both conditions (N = 35 per condition). The treatment condition significantly outperformed the control in three individual questions. Question 1 improved dramatically from 2.9\% to 51.4\% correct (p < 0.001), Question 3 from 11.4\% to 48.6\% (p < 0.001), and Question 2 from 17.1\% to 40.0\% (p = 0.026). The pattern of improvements suggests that \system is particularly effective at helping users track specific editorial changes and AI contributions. Question 1's dramatic improvement (from 2.9\% to 51.4\%) indicates that our visualization successfully addresses a fundamental challenge in understanding AI-assisted writing: identifying where changes occurred. The substantial gains in Questions 2 and 3, which focus on understanding AI feedback integration, suggest that making the revision process visible helps users develop better mental models of human-AI collaboration. However, the relatively low baseline performance across all questions (2.9\%--17.1\%) confirms that without appropriate tools, readers struggle significantly to understand AI's role in collaborative writing, making transparency tools like DraftMarks essential rather than merely helpful. In general, participants in the treatment group scored significantly higher than controls ($\mu = 4.29$ vs. $\mu = 2.86$ out of 7, $U = 335.5$, $p < 0.001$, $r = 0.52$), representing a 50\% improvement with a large effect size.

\subsubsection{Cognitive Load Survey.} Based on our analysis of the CLS responses, participants reported moderate intrinsic load ($\mu = 4.31$, $\sigma = 2.66$) and low extraneous load ($\mu = 3.03$, $\sigma = 2.46$) while using the system. Furthermore, their rated self-perceived learning, captured by germane load, was high ($\mu = 7.42$, $\sigma = 1.96$). This cognitive load profile suggests that \system successfully balances information richness with cognitive accessibility. The moderate intrinsic load indicates that understanding AI-assisted writing processes requires meaningful mental effort, which is appropriate given the complexity of human-AI collaboration. The high germane load with relatively low variability ($\sigma = 1.96$) suggests consistent learning experiences across participants, indicating that the system effectively supports comprehension regardless of individual differences in technical background or reading strategies.

\subsubsection{System Usability Scale.} \system achieved a strong average SUS score of 80.5 ($\sigma = 12.8$), placing it well above the 68 threshold for acceptable usability. Analysis of individual SUS items reveals that users found the system easy to use (SUS 3: 74\% agreement) and expressed confidence in using it (SUS 9: 77\% agreement). However, some participants indicated they would need technical support to use the system (SUS 4: 69\% disagreement with needing support) and found various functions well integrated (SUS 5: 89\% agreement). The strong SUS score is particularly meaningful given that participants used \system with no prior training, suggesting the visualization design successfully leverages familiar reading and annotation paradigms. The high agreement on function integration (SUS 5: 89\%) indicates that our approach of embedding transparency features directly into the document interface feels natural to users. However, the mixed responses on technical support needs (SUS 4) suggest that while most users can operate the system independently, the complexity of tracking AI contributions may initially overwhelm some users. 

\subsubsection{Transparency Measure.} Analysing the three transparency questions, we notice participants in both conditions reported relatively high transparency perceptions with the \system condition consistently outperformed AI-Highlight marginally across all. For \textit{confidence in assessment}, DraftMarks participants scored higher ($\mu = 5.81$, $\sigma = 1.00$) compared to AI-Highlight ($\mu = 5.27$, $\sigma = 1.26$). Similarly, for \textit{information sufficiency}, DraftMarks again showed superior performance ($\mu = 5.74$, $\sigma = 1.32$) versus AI-Highlight ($\mu = 5.24$, $\sigma = 1.09$). The \textit{perception of human effort visibility} followed the same pattern, with DraftMarks ($\mu = 5.93$, $\sigma = 1.04$) outperforming AI-Highlight ($\mu = 5.48$, $\sigma = 1.18$). Despite these modest differences in transparency perceptions, participants in the AI-Highlight condition performed significantly worse on comprehension measures, suggesting that surface-level visibility of AI assistance (through chat logs and highlighted text) created an illusion of process understanding without meaningful insight into the writing workflow itself.

\subsection{Qualitative Feedback}
The open-ended responses revealed distinct patterns between conditions regarding what helped users understand human-AI collaboration. For \system condition 74\% (26/35) participants shared their thoughts for F1 and F2, while for AI-Highlighted 91\% (31/35) responded. For F1, over half of \system respondents (15/26, 57\%) praised eraser crumbs, residual glue and masking tape for helping them see what stuck in the writing from AI and what didn't. In contrast, 31\% of AI-Highlight participants (10/32) experienced confusion about attribution, with one participant noting \textit{``[It was] hard to tell how much help the AI gave or where the human made changes''}. While only 8\% of system users (2/26) experiencing confusing, one respondent requested clearer legends for the process cues. The 35\% of AI-Highlight users who struggled did so with fundamental questions of \textit{``figuring out exactly which parts were written by the AI and which by the human.''}
\section{Discussion and Future Work}
Based on our design and evaluation, here we reflect on the implications of \system for readers and writers. Based on our reflection, we identify tensions and opportunities that warrant further investigation.

\subsection{The Cost of Making the Writing Process Legible}
Writers have long used signaling as a \textit{deliberate} rhetorical strategy to shape the way their work is read~\cite{lorch1989text}. Titles, headings, and typographic emphasis such as \textbf{bold} and \textit{italicized} text help readers infer importance, intent, and argumentative flow~\cite{lorch1989text}. These signals are authored under the writer's control and serve their communication goals. In contrast, \system introduces a form of implicit signaling: cues that are automatically generated from interaction data and surfaced within the text to make the writing process legible. The result is a document that not only communicates ideas but also reveals how those ideas were constructed, who revised, who delegated, and how much cognitive effort each section reflects. This form of visibility offers clear benefits. It supports more transparent forms of authorship, helps readers assess originality and deliberation, and enables richer feedback conversations between writers, instructors, collaborators, or reviewers. 

However, there are other trade-offs to consider when making the process implicitly visible. Because these signals are not author-curated, they may surface aspects of writing that feel personal, incomplete, or even compromising, such as moments of hesitation, minimal revision, or heavy reliance on AI-generated content. With \system, what was once invisible now becomes part of the public artifact. In an educational and evaluative setting, this visibility can redefine the way effort, originality, and ability are perceived. The core trade-off is between \textit{interpretive transparency} and \textit{authorial agency}, i.e., helping readers understand how text was constructed and giving writers control over what is disclosed. This raises important design questions for \system that need to be addressed in future work. Should writers be able to redact or annotate process signals? Can visibility be audience-specific, e.g., shown to instructors but hidden in publication? And how might authors be supported in interpreting and managing their own signals before others? Questions about who decides what becomes visible, to whom, and when, are important topics for future exploration. 

\subsection{Supporting Writer Reflection}
Although \system is designed to help readers interpret AI-assisted writing, it has the potential to offer direct benefits to writers themselves. In traditional pen-and-paper writing, the act of drafting leaves behind a trail that helps writers reflect on their processes and decisions. In contrast, AI-generated text is often presented to writers fully formed and blended within the context of the current text, making it easy to accept suggestions without revisiting or refining them. This can potentially erode opportunities for deliberate thinking, reflection, and self-assessment. \system can alleviate this concern by externalizing writers' interactions with AI and revealing where their effort has (or has not) been directed. Visual cues such as eraser marks, prompt iteration indicators, and masking tape overlays make the invisible parts of their process tangible. This visibility can encourage writers to pause and reflect on questions such as \textit{ Have I done enough to make this idea my own? Am I critically engaged with AI suggestions, or simply accepting them?} By surfacing patterns of revision, delegation, and iteration, DraftMarks helps writers monitor their writing habits and prompts more intentional engagement with the text. This type of metacognitive support is especially beneficial in educational settings. For students, \system can promote awareness of their development as writers, providing a concrete record of revision that supports growth and feedback conversations (e.g.,~\cite{Guo2024FromPT}). Future work can explore specific interactive affordances and guidance for writers when co-authoring with AI. 

Our approach also shares conceptual grounding with linkography~\cite{goldschmidt2014linkography}, which visualizes how design ideas develop over time by tracing connections between moves. In this vein, techniques such as \system could evolve towards ``forward-looking'' linkography, visualizing emerging conceptual connections and gaps in human-ai collaborative writing. A potential direction for future work exploring generative linkography to support collaborative writing and enable new forms of epistemic feedback.

\subsection{Skeuomorphic Signals and Interpretation}
Skeuomorphic visual metaphors, as a design choice, come with both strengths and limitations. Skeuomorphism can leverage familiar physical affordances to make abstract processes like human-AI interaction visually intuitive and contextually embedded~\cite{Lee:2022:BiosignalVisVR}. For writers and readers accustomed to pencil-and-paper revision practices, these metaphors evoke meaningful analogies, such as eraser marks suggesting revision and uncertainty. Compared to alternative techniques for process visibility (see Section~\ref{sec:related_work}), \systems strength is that it integrates process cues into the text itself over detached metadata or summary metrics.  However, these cues are not universally legible. For writers primarily exposed to digital environments or in contexts where physical revision artifacts are less common, skeuomorphic signals can lack resonance or even add confusion~\cite{Urbano:2022:SkeuomorphismFlatDesign}. Future work can explore other metaphors within the space of typography, such as new fonts, the use of dynamic underlines, and color gradients, to represent process data in a more culturally neutral or scalable form. However, much like the floppy disk icon that is still widely used to represent ``save'' functionality, despite many users never having seen or used one. Skeuomorphic cues can persist even as their original referents lose relevance, raising questions about the intuitive nature of such metaphors across generations and cultures.

\subsection{Stakeholder-Specific Design Considerations}
Our formative evaluation revealed striking differences in stakeholder needs for process transparency. Teachers valued detailed collaboration insights for formative assessment, while academic reviewers viewed process visualization as unnecessary ``extra work'' that detracted from their efficiency-focused evaluation goals. General readers fell between these extremes, using transparency signals to assess authenticity and decide whether content warranted their attention. These findings suggest \system requires stakeholder-specific modes rather than a universal interface. An ``instructor mode'' might expose comprehensive process data, while a ``publication mode'' could focus on high-level authenticity signals. To realize this, future work should investigate configurable controller algorithms that flexibly govern which dimensions of transparency are emphasized, suppressed, or abstracted for different audiences. However, this raises questions about who controls visibility settings when documents serve multiple audiences simultaneously. Designing flexible but principled controller algorithms thus requires not only technical advances in interface adaptability but also \textbf{normative frameworks} to arbitrate control, protect against selective disclosure, and uphold shared standards of writer agency and authenticity~\cite{Birnholtz2012TrackingCI, Aburass2024AuthenticityIA, Alfassi2025FanfictionIT}.

\subsection{Limitations of Process Transparency: The Missing Citation Layer}
While \system surfaces human–AI collaboration patterns, it does not address AI training data provenance. Our tool reveals when AI suggestions were integrated into the writing process, but it cannot expose the specific sources that informed those AI generations. This limitation is significant given growing concerns about AI systems reproducing copyrighted content without attribution, either verbatim or through close paraphrase. Such outputs risk crossing the line into plagiarism, especially when users unknowingly incorporate AI-generated text that echoes protected works~\cite{huang2023citation}. Beyond individual academic integrity, this raises broader copyright challenges: without visibility into provenance, it is impossible to determine whether AI contributions constitute fair use, derivative works, or potential infringement. Future process-visualization tools should therefore not only trace collaboration patterns but also provide mechanisms for AI citation transparency, making visible when generated text may carry obligations of attribution. Realizing this vision, however, requires advances in AI explainability and provenance-tracking that extend well beyond current technical capabilities~\cite{gao2023enabling, huang2025authorship}.
\section{Conclusion}
As AI becomes a coauthor in everyday writing, understanding the creative processes by which a text was constructed has become essential for meaningful interpretation, evaluation, and learning. \system introduces a new form of process-aware signaling by embedding skeuomorphic visual cues directly into the text, helping readers see where and how AI was involved and where human effort was concentrated. By shifting process metadata from the margins to the foreground, \system provides transparency, encourages writer reflection, and can facilitate more grounded feedback from educators or peers. Our evaluation demonstrates \systems affordances in helping readers better understand and assess human effort in writing, showing that in-text visualizations can meaningfully enhance how AI-assisted work is interpreted across educational and academic contexts.

\bibliographystyle{ACM-Reference-Format}
\bibliography{99_refs}

\end{document}